\documentclass[traditabstract,longauth]{aa}
\usepackage{graphicx}
\usepackage{xcolor,soul}
\usepackage{svg}
\usepackage{longtable}
\usepackage{array}
\usepackage{lscape}
\usepackage{adjustbox}
\usepackage{xcolor}
\usepackage[toc,page]{appendix}

\usepackage{txfonts}
\usepackage{amsmath,siunitx}

\usepackage[breaklinks, colorlinks, citecolor=blue, linkcolor=blue]{hyperref}

\usepackage{xcolor}

\usepackage{orcidlink}
\usepackage{hyperref}

\providecommand{\mst}{\ensuremath{\,{\rm M_\odot}}}
\providecommand{\rst}{\ensuremath{\,{\rm R_\odot}}}

\providecommand{\arcsec}{$^{\prime \prime}$}

\graphicspath{{./}{figures/}}

\begin{document}

   \title{Unraveling the Brown Dwarf Desert: Four New Discoveries and a Unifying, Period-Coded Picture\thanks{}}

   \subtitle{}

   \author{J\'an \v{S}ubjak 
          \inst{1,2}\orcidlink{0000-0002-5313-9722}
          \and
          Rafael Brahm
          \inst{3,4}\orcidlink{0000-0002-9158-7315}
          \and
          Jozef Lipt\'{a}k
          \inst{1,5}
          \and
          Jan Eberhardt
          \inst{6}\orcidlink{0000-0003-3130-2768}
          \and
          Marcelo Tala Pinto
          \inst{7}\orcidlink{0009-0004-8891-4057}
          \and
          Sarah L. Casewell
          \inst{8}\orcidlink{0000-0003-2478-0120}
          \and
          Thomas Henning
          \inst{6}\orcidlink{0000-0002-1493-300X}
          \and
          Katharine Hesse
          \inst{9}\orcidlink{0000-0002-2135-9018}
          \and
          Trifon Trifonov
          \inst{6,10,11}\orcidlink{0000-0002-0236-775X}
          \and
          Andr\'es Jord\'an
          \inst{3,4}\orcidlink{0000-0002-5389-3944}
          \and
          Felipe I. Rojas
          \inst{12}\orcidlink{0000-0003-3047-6272}
          \and
          Michaela V\'{i}tkov\'{a}
          \inst{1,13}\orcidlink{0000-0002-2994-2929}
          \and
          Helem Salinas
          \inst{3}\orcidlink{0009-0009-8795-4563}
          \and
          Gavin Boyle
          \inst{14,15}\orcidlink{0009-0009-2966-7507}
          \and
          Vincent Suc
          \inst{3,4,14}\orcidlink{0000-0001-7070-3842}
          \and
          Luca Antonucci
          \inst{16}\orcidlink{0009-0003-4135-1296}
          \and
          Krzysztof Bernacki
          \inst{17}\orcidlink{0000-0003-4647-7114}
          \and
          C\'{e}sar Brice\~{n}o
          \inst{18}\orcidlink{0000-0001-7124-4094}
          \and
          Karen A. Collins
          \inst{2}\orcidlink{0000-0001-6588-9574}
           \and
          Jorge Fern\'andez Fern\'andez
          \inst{19,20}\orcidlink{0000-0002-1416-2188}
          \and
          Samuel Gill
          \inst{19,20}\orcidlink{0000-0002-4259-0155}
          \and
          Jan Jan\'ik
          \inst{13}\orcidlink{0000-0002-6384-0184}
          \and
          Nicholas Law
          \inst{21}\orcidlink{0000-0001-9380-6457}
          \and
          Andrew W. Mann
          \inst{21}\orcidlink{0000-0003-3654-1602}
          \and
          James McCormac
          \inst{19,20}\orcidlink{0000-0003-1631-4170}
          \and
          Adam Popowicz
          \inst{21}\orcidlink{0000-0003-3184-5228}
          \and
          Daniel Sebastian
          \inst{22}\orcidlink{0000-0002-2214-9258}
          \and
          Marek Skarka
          \inst{1}\orcidlink{0000-0002-7602-0046}
          \and
          J\'an V\'aclav\'ik
          \inst{23}
          \and
          Leonardo Vanzi
          \inst{16}
          \and
          Richard G. West
          \inst{19,20}\orcidlink{0000-0001-6604-5533}
          \and
          Francis P. Wilkin
          \inst{24}\orcidlink{0000-0003-2127-8952}
          \and
          Carl Ziegler
          \inst{25}\orcidlink{0000-0002-0619-7639}
          }

   \institute{Astronomical Institute, Czech Academy of Sciences, Fri{\v c}ova 298, 251 65, Ond\v{r}ejov, Czech Republic
         \and
         Center for Astrophysics ${\rm \mid}$ Harvard {\rm \&} Smithsonian, 60 Garden Street, Cambridge, MA 02138, USA
         \and
         Facultad de Ingenier\'ia y Ciencias, Universidad Adolfo Ib\'{a}\~{n}ez, Av. Diagonal las Torres 2640, Pe\~{n}alol\'{e}n, Santiago, Chile
         \and
         Millennium Institute for Astrophysics, Nuncio Monse\~{n}or Sotero Sanz 100, Of. 104, Providencia, Santiago, Chile
         \and
         Astronomical Institute of Charles University, V Hole\v{s}ovi\v{c}k\'{a}ch 2, CZ-180 00 Prague, Czech Republic
         \and
         Max-Planck-Institut für Astronomie, Königstuhl 17, D-69117 Heidelberg, Germany
         \and
         Department of Astronomy, The Ohio State University, 140 W. 18th Avenue, Columbus, OH 43210, USA
         \and
         School of Physics and Astronomy, University of Leicester, University Road, Leicester LE1 7RH, UK
         \and
         Kavli Institute for Astrophysics and Space Research, Massachusetts Institute of Technology, 70 Vassar St., Cambridge, MA 02139, USA
         \and
         Department of Astronomy, Sofia University ``St Kliment Ohridski'', 5 James Bourchier Blvd, BG-1164 Sofia, Bulgaria
         \and
         Landessternwarte, Zentrum f\"ur Astronomie der Universit\"at Heidelberg, K\"onigstuhl 12, D-69117 Heidelberg, Germany
         \and
         Instituto de Astrof\'isica, Pontificia Universidad Cat\'olica de Chile, Av. Vicu\~na Mackenna 4860, 7820436 Macul, Santiago, Chile
         \and
         Department of Theoretical Physics and Astrophysics, Faculty of Science, Masaryk University, Kotl\'{a}\v{r}sk\'{a} 2, CZ-61137 Brno, Czech Republic
         \and
         El Sauce Observatory --- Obstech, Coquimbo, Chile
         \and
         Cavendish Laboratory, J. J. Thomson Avenue, Cambridge, CB3 0HE, UK
         \and
         Department of Electrical Engineering and Center of Astro Engineering, Pontificia Universidad Cat\'olica de Chile, Av. Vicu\~na Mackenna 4860, 7820436 Macul, Santiago, Chile
         \and
         Silesian University of Technology, Akademicka 16, 44-100 Gliwice, Poland
         \and
         Cerro Tololo Inter-American Observatory, Casilla 603, La Serena, Chile
         \and
         Department of Physics, University of Warwick, Gibbet Hill Road, Coventry CV4 7AL, UK
         \and
         Centre for Exoplanets and Habitability, University of Warwick, Gibbet Hill Road, Coventry CV4 7AL, UK
         \and
         Department of Physics and Astronomy, The University of North Carolina at Chapel Hill, Chapel Hill, NC 27599-3255, USA
         \and
         Thueringer Landessternwarte Tautenburg, Sternwarte 5, 07778 Tautenburg, Germany
         \and
         Institute of Plasma Physics of the Czech Academy of Sciences, Research Centre for Special Optics and Optoelectronic Systems TOPTEC, U Slovanky 2525/1a, 182 00, Prague, Czech Republic
         \and
         Department of Physics and Astronomy, Union College, 807 Union St., Schenectady, NY 12308, USA
         \and
         Dunlap Institute for Astronomy and Astrophysics, University of Toronto, 50 St. George Street, Toronto, Ontario M5S 3H4, Canada
        }

   \date{\today{}; \today{}}

 
\abstract{We present four newly validated transiting brown dwarfs identified through \textit{TESS} photometry and confirmed with high-precision radial velocity measurements obtained from the FEROS and PLATOSpec spectrographs. Notably, three of these companions exhibit orbital periods exceeding 100 days, thereby expanding the sample of long-period transiting brown dwarfs from two to five systems. The host stars of long-period brown dwarfs show mild subsolar metallicity. These discoveries highlight the expansion of the metal-poor, long-period distribution and help us better understand the brown dwarf desert. In our comparative analysis of eccentricity and metallicity demographics, we utilize catalogues of long-period giant planets, brown dwarfs, and low-mass stellar companions. After accounting for tidal influences, the eccentricity distribution aligns with that of low-mass stellar binaries, presenting a different profile than that observed within the giant planet population. Additionally, the metallicity of the host stars reveals a noteworthy trend: short-period transiting brown dwarfs are predominantly associated with metal-rich stars, whereas long-period brown dwarfs are more often found around metal-poor stars, demonstrating statistical similarities to low-mass stellar hosts. This trend has also been previously observed in studies of hot and cold Jupiters and points to a period-coded mixture of channels. A natural explanation is that most brown dwarfs originate from fragmentation at wider separations, with long-period systems retaining this stellar-like imprint, while only those embedded in massive, long-lived, metal-rich protoplanetary discs are efficiently delivered and stabilised to short orbits.}

\keywords{planetary systems -- techniques: photometric -- techniques: spectroscopic -- techniques: radial velocities -- planets and satellites: formation -- planets and satellites: dynamical evolution and stability
               }
\maketitle


\section{Introduction}\label{sec:introduction}

Brown dwarfs (BDs) occupy the mass range between giant planets and low-mass stars. The deuterium-burning limit, which is dependent on composition, is situated near approximately $\sim$13~$M_{\rm Jup}$, while the hydrogen-burning threshold is around $\sim$0.075~$M_\odot$, equivalent to approximately $78-80~M_{\rm Jup}$ \citep{Baraffe2002, Spiegel2011}. Initial radial-velocity (RV) surveys have revealed a scarcity of close BD companions within a few astronomical units from FGK stars, commonly referred to as the "brown-dwarf desert". This observation has prompted a deeper investigation into whether BDs from the desert form and migrate in a manner similar to giant planets or if they exhibit behavior akin to stellar binaries. Addressing this inquiry would provide critical insights into the formation history, which is regarded as the most fundamental distinguishing factor \citep{burrows01}. According to this perspective, substellar objects that form through disk instability, akin to stars, would be classified as brown dwarfs, whereas those that form via core accretion would be identified as planets. However, determining the formation history of individual systems through observational means presents considerable challenges; thus, studying the demographics of the entire population may offer a better opportunity to obtain conclusive answers.

Over the past decade, the number of \emph{transiting} brown dwarfs has increased substantially, largely thanks to \textit{Kepler}/K2 and particularly \textit{TESS} \citep[e.g.,][]{Persson2019,Subjak2020,Carmichael2022,Subjak2024,Vowell2025}. Transiting systems offer model-independent measurements of radii and dynamical masses, thereby addressing the complexities associated with mass–radius–age degeneracies and eliminating the sin(i) ambiguity that often complicates radial velocity (RV) detections. With the current sample of more than 50 transiting brown dwarfs, it becomes feasible to investigate population-level trends with a reduced number of selection biases compared to earlier RV studies \citep{Vowell2025}.

A long-standing hypothesis suggests that transiting BDs may consist of two populations, differentiated around $\sim$40–45~$M_{\rm Jup}$ \citep{Ma2014}. Higher-mass objects in this category exhibit binary-like eccentricities, whereas lower-mass counterparts resemble giant planets. However, a fresh perspective emerging from the expanding census of transiting BDs does not support evidence for a distinct transition in the masses \citep{Vowell2025}. Hence, the question of whether demographic divides exist within the transiting BD population—and, importantly, whether mass, eccentricity, metalicity, or orbital period serves as the more significant organizing variable—remains unresolved.

Two recent observational developments now make this question timely. The sample is beginning to populate a \emph{long-period tail} of transiting brown dwarfs that was previously underrepresented in transit work \citep{Carmichael21,Henderson24,A_New_BD_Mstar_2025}. These BDs are crucial leverage: they sit beyond the most intense tidal regime, retain more natal orbital memory, and provide a cleaner stage for testing whether BD demographics align with giant planets or with stellar companions.

In this paper we present the discovery and validation of four new transiting BDs from \textit{TESS} and follow-up spectroscopy, including three long-periodic BDs with periods larger than 100 days, which substantially extend the long-period tail (only two such systems were known). The paper includes a description of the observations in Section \ref{sec:observations}, data analysis in Section \ref{sec:analysis}, a discussion in Section \ref{sec:discussion}, and a summary of our results in Section \ref{sec:summary}.


%
%

\begin{table*}
	\centering
	\caption{System parameters for stars in our sample.}
	\label{tab1}
	\scalebox{0.81}{
	\begin{tabular}{lcccccc} 
		\hline
		\hline
		System       & TIC\,9344899 & TIC\,52059926 & TIC\,13344668 & TIC\,63921468 & Source\\
		\hline
        RA$_{J2000}$ (hh:mm:ss.ss) & 04 51 10.54 & 01 07 05.44 & 05 14 52.75 & 00 53 05.00 & 1\\
        Dec$_{J2000}$ (d:':") & -08 26 50.35 & -68 22 05.17 & -29 43 46.78 & -24 57 06.34 & 1\\
        \smallskip\\
        TESS $T$ mag & $12.735 \pm 0.006$ & $12.730 \pm 0.006$ & $12.673 \pm 0.006$ & $8.879 \pm 0.007$ & 2\\
        $Gaia$ $G$ mag & $13.365 \pm 0.003$ & $13.413 \pm 0.003$ & $13.365 \pm 0.003$ & $9.207 \pm 0.003$ & 1\\
        Tycho $B_T$ mag & $14.037 \pm 0.011$ & $14.934 \pm 0.017$ & $14.876 \pm 0.023$ & $9.818 \pm 0.025$ & 3\\
        Tycho $V_T$ mag & $13.291 \pm 0.092$ & $13.724 \pm 0.092$ & $13.700 \pm 0.126$ & $9.344 \pm 0.020$ & 3\\
        2MASS $J$ mag & $12.118 \pm 0.027$ & $11.747 \pm 0.024$ & $11.670 \pm 0.022$ & $8.423 \pm 0.020$ & 4\\
        2MASS $H$ mag & $11.829 \pm 0.026$ & $11.133 \pm 0.022$ & $11.095 \pm 0.024$ & $8.207 \pm 0.029$ & 4\\
        2MASS $K_S$ mag & $11.759 \pm 0.034$ & $11.023 \pm 0.021$ & $10.971 \pm 0.023$ & $8.182 \pm 0.033$ & 4\\
        WISE1 mag & $11.675 \pm 0.022$ & $10.990 \pm 0.023$ & $10.914 \pm 0.023$ & $8.131 \pm 0.022$ & 5\\
        WISE2 mag & $11.735 \pm 0.021$ & $11.077 \pm 0.021$ & $10.966 \pm 0.021$ & $8.143 \pm 0.020$ & 5\\
        WISE3 mag & $11.381 \pm 0.212$ & $11.047 \pm 0.109$ & $10.841 \pm 0.096$ & $8.157 \pm 0.021$ & 5\\
        WISE4 mag & $8.484 \pm 0.250$ & $9.112 \pm 0.250$ & $9.200 \pm 0.250$ & $7.800 \pm 0.209$ & 5\\
        \smallskip\\
	    $\mu_\alpha\,cos(\delta)$ (mas/yr) & $9.458 \pm 0.013$ & $-19.323 \pm 0.012$ & $4.073 \pm 0.011$ & $23.525 \pm 0.022$ & 1\\
	    $\mu_{\delta}$ (mas/yr) & $4.227 \pm 0.011$ & $-11.843 \pm 0.013$ & $68.687 \pm 0.013$ & $-19.150 \pm 0.020$ & 1\\
	    Parallax (mas) & $1.894 \pm 0.013$ & $4.586 \pm 0.011$ & $5.064 \pm 0.012$ & $4.201 \pm 0.021$ & 1\\
		\hline
		\hline
		
	\end{tabular}
	}
	\smallskip\\
References: 1 - $Gaia$ DR3, \citet{Gaia21}; \\ 2 - TESS, \citet{Stassun18}; 3 - Tycho, \citet{Hog00}; 4 - 2MASS, \citet{Cutri03}; 5 - WISE, \citet{Wright10} \\
\end{table*}

\begin{table}
\caption{Dates of TESS observations for our sample of stars.}             
\label{table:tess_o_d}      
\centering                          
\begin{tabular}{c c c c}        
\hline\hline                 
Star & Sector & Cadence (s) & Source \\    
\hline                        
   TIC\,9344899 & 5 & 1800 & QLP \\
   & 32 & 600 & QLP \\
   \hline
   TIC\,52059926 & 1 & 1800 & SPOC \\
    & 2 & 1800 & SPOC \\
    & 28 & 600 & SPOC \\
    & 29 & 600 & SPOC \\
    & 68 & 120 & SPOC \\
    & 69 & 120 & SPOC \\
    \hline
   TIC\,13344668 & 5 & 1800 & SPOC \\
    & 32 & 20 & SPOC \\
    \hline
   TIC\,63921468 & 3 & 1800 & SPOC \\
    & 30 & 120 & SPOC \\
    \hline
    
\hline                                   
\hline
\end{tabular}
\end{table}

\section{Observations}\label{sec:observations}

\subsection{TESS light curves}\label{sec:TESS}

TIC 9344899, TIC 52059926, TIC 13344668, and TIC 63921468 were first imaged by the Transiting Exoplanet Survey Satellite \citep[TESS;][]{Ricker15} in long-cadence Full Frame Image (FFI) mode at a 1800-second cadence during the Prime Mission (PM) and re-observed at shorter cadences during the first and second Extended Missions (EMs). The sectors of observation and cadences for each are detailed in Table \ref{table:tess_o_d}. With the available photometry, two planet candidates were alerted as TESS Objects of Interest \citep[TOIs;][]{Guerrero21} by the TESS Science Office -- a single transit planet candidate promoted through the CTOI (community TOI) review process from Rafael Brahm's CTOI contribution designated as TOI 2530.01 (TIC\,52059926) and a planet candidate with an orbital period of 8.58 days identified through the TESS Faint Star Search \citep{Kunimoto2022} leveraging data from the MIT Quick Look Pipeline \citep[QLP;][]{Huang2020} was alerted as TOI 4701.01 (TIC\,9344899). The remaining two stars were each designated as the hosts of single transit CTOIs on the ExoFOP (Exoplanet Follow-up Observing Program\footnote{\url{https://exofop.ipac.caltech.edu/tess/}}, with TIC 13344668.01 submitted by the WINE team and TIC 63921468.01 detected by Helem Salinas \citep{Salinas25} though neither have been promoted to TOIs to date.



The light curves (LCs) for TIC\,52059926, TIC\,13344668, and TIC\,63921468 were retrieved from the Mikulski Archive for Space Telescopes (MAST)\footnote{\url{https://mast.stsci.edu/portal/Mashup/Clients/Mast/Portal.html}} using the {\tt lightkurve} software package \citep{Lightkurve18}. We specifically employed the Pre-search Data Conditioning Simple Aperture Photometry (PDCSAP) LCs \citep{Smith12, Stumpe2012, Stumpe2014}, processed through the SPOC pipeline, which is designed to correct for systematic instrumental artifacts. To analyze crowding effects, pixel response functions (PRFs) were utilized, and a correction for crowding was incorporated into the PDCSAP flux time series.

For TIC\,9344899, we analyzed the spline-detrended KSPSAP light curves generated by the QLP from the TESS full-frame images (FFIs). A comprehensive overview of the QLP can be found in the works of \citet{Huang20,Fausnaugh2020,kunimoto2021}. A summary of the TESS data utilized in our analysis for each target is provided in Table \ref{table:tess_o_d}.

To validate the QLP LCs we have also extracted a TESS FFI LCs for TIC\,9344899 using \texttt{TESSCut} \citep{Castro-Gonzalez24} and the \texttt{lightkurve} package. Pixel-level decorrelation (PLD) was performed with four PCA components and a cubic spline. After PLD, we applied a low-order flattening ($\approx$1-day window; polynomial order $=2$) that preserved in-transit data, and normalized the result. Fitting the phase-folded light curve yielded an \emph{observed} transit depth of $\delta_{\mathrm{obs}} = 3.58$~ppt from this FFI extraction. To quantify dilution from nearby stars in the large TESS pixels, we ran \texttt{tess-cont} \citep{Castro-Gonzalez24} with \textit{Gaia}~DR3 sources and the PRF ``accurate'' option, using an aperture matched to the one above. The PRF model returned $\mathrm{CROWDSAP} \simeq 0.51$ (i.e., $\sim$49\% of the flux inside the aperture is from contaminating stars). Most of the contamination is coming from the source Gaia DR3 3186620924792864512, which has the same magnitude as TIC\,9344899, and a separation of about $30\arcsec$. We therefore corrected the measured depth via
\begin{equation}
\delta_{\mathrm{true}} = \frac{\delta_{\mathrm{obs}}}{\mathrm{CROWDSAP}} \approx \frac{3.58~\mathrm{ppt}}{0.51} \approx 7~\mathrm{ppt},
\end{equation}
which is not far from the depth obtained from the QLP light curve ($8.6$~ppt) given the different apertures and cotrending treatments (CROWDSAP of 0.43 would lead to the given depth). Independent ground-based photometry with the SUTO-UZPW50 0.5\,m telescope on 2022-11-17 in the $R$ band detected a $6.7$\,ppt transit using an aperture radius of $7.04^{\prime\prime}$ that is small enough to exclude the $\sim30^{\prime\prime}$ neighbor. This uncontaminated depth agrees with the deblended TESS value to within $\sim5\%$, consistent with modest bandpass/limb-darkening differences and normal systematics. In all subsequent modeling, we use the TESS FFI LCs corrected for dilution.

We employed the Python package {\tt citlalicue} \citep{Barragan22} to mitigate residual stellar variability in the light curves (LCs). By integrating Gaussian process (GP) regression with transit models generated through the {\tt pytransit} framework \citep{Parviainen15}, {\tt citlalicue} produced a composite model that accounted for both light curve variability and transit signals. We subsequently subtracted the variability component, yielding flattened LCs that isolate the transit-related photometric variations. The light curves, before and after this processing, are illustrated in Fig.~\ref{fig:tess_lc}.

To assess the impact of nearby sources on the TESS photometry, we compared the {\it Gaia} DR2/DR3 catalogs with the TESS target pixel file (TPF) using the {\tt TESS-cont} tool \citep{Castro-Gonzalez24}. This tool identifies stars that may dilute the TESS signal. In Figure \ref{fig:field_image}, we present a heatmap indicating the percentage of flux from the target star contained in each pixel. Our analysis shows that nearby sources contribute 6.5\% (TIC\,52059926), 0.7\% (TIC\,13344668), and 0.1\% (TIC\,63921468) of the total flux, suggesting that the contamination is minimal and has been appropriately corrected in the PDCSAP.

\subsection{Ground-based Photometric Follow-up}

As part of the TFOP, ground-based photometric data were collected for TIC\,9344899, and TIC\,13344668. The observations were scheduled using Transit Finder, a customized version of the {\tt Tapir software} \citep{Jensen13}, and the photometric data were extracted using {\tt AstroImageJ} \citep{Collins17}. All light curve data are available on the {\tt EXOFOP-TESS} website.\footnote{\href{https://exofop.ipac.caltech.edu/tess/}{https://exofop.ipac.caltech.edu/tess/}}

One full transit of TIC\,9344899 was observed on 2022 November 17 in the R band with the Silesian University of Technology (SUTO) 0.5m telescope. The SUTO telescope is equipped with a 9576 × 6388 pixel QHY600 camera with an image scale of $0.88\arcsec$ pixel$^{-1}$. The aperture radius was 8 pixels. We used the photometry in our global modeling of the system.

The Observatoire Moana (OM) is a global network of small-aperture robotic telescopes. We utilized two OM stations located at the El Sauce Observatory in Chile (OMES500 and OMES600) and one station at the Siding Spring Observatory in Australia (OMSSO500) to observe three transits of TIC 9344899. OMES600 is a 0.6-meter corrected Dall-Kirkham robotic telescope equipped with an Andor iKon-L 936 deep depletion 2k×2k CCD, which has a scale of 0.67 arcseconds per pixel. OMES500 is a 0.5-meter Riccardi Dall-Kirkham telescope also equipped with an Andor iKon-L 936 deep depletion 2k×2k CCD, featuring a scale of 0.7 arcseconds per pixel. OMSSO500 consists of a 0.5-meter RCOS Ritchey-Chretien telescope paired with an FLI ML16803 4k×4k CCD that has a pixel scale of 0.47 arcseconds per pixel, operating with 2×2 binning. On September 14, 2023, we used OMSSO500 to observe the egress of transit. On November 22, 2023, we employed OMES600 to observe a full transit. Finally, on November 25, 2024, we utilized OMES500 to observe another full transit. The data from OMES is automatically processed to create light curves using differential photometry through dedicated pipelines \citep{toi2525, brahm:2023}.

The Next Generation Transit Survey operates a collection of twelve robotic telescopes situated at Cerro Paranal, Chile. Each telescope features a 20 cm aperture and is equipped with back-illuminated deep-depletion CCD cameras \citep{Wheatley2018}. The raw images were processed following the method discussed in \citet{Wheatley2018}. Partial transits of TIC\,9344899\,b were observed on the following dates: October 12, 2024; November 07, 2024; and January 06, 2025. In addition, TIC\,13344668 was monitored nightly (whenever weather permitted) between August 17, 2017, and March 28, 2018. During this period, two transits occurred; however, they were not observed because they happened either during the day or in poor weather conditions.

Our team conducted additional photometry using the Las Cumbres Observatory Global Telescope \citep[LCOGT;][]{Brown13} 1.0-m network. The telescopes are equipped with 4096×4096 pixel SINISTRO cameras that have an image scale of 0.389\,arcsec per pixel. One transit was observed in the Sloan $i'$ filter on November 12, 2025. The images were calibrated using the standard {\tt LCOGT BANZAI} pipeline \citep{McCully18}. The aperture radius covers the range from 5.5 to 10.5\,arcsecs.

Our team conducted nightly (whenever weather permitted) NGTS monitoring of TIC\,13344668 between August 17, 2017, and March 28, 2018. 

\subsection{Contamination from nearby sources}\label{sec:ao_image}

To eliminate any potential sources of dilution within the {\it Gaia} separation limit of 0.4\arcsec, we obtained high-resolution images using adaptive optics and speckle imaging of TIC\,52059926. This was done as part of follow-up observations that were coordinated through the TESS Follow-up Observing Program (TFOP) High-Resolution Imaging Sub-Group 3 (SG3).

\begin{figure}
\centering
\includegraphics[width=0.45\textwidth]{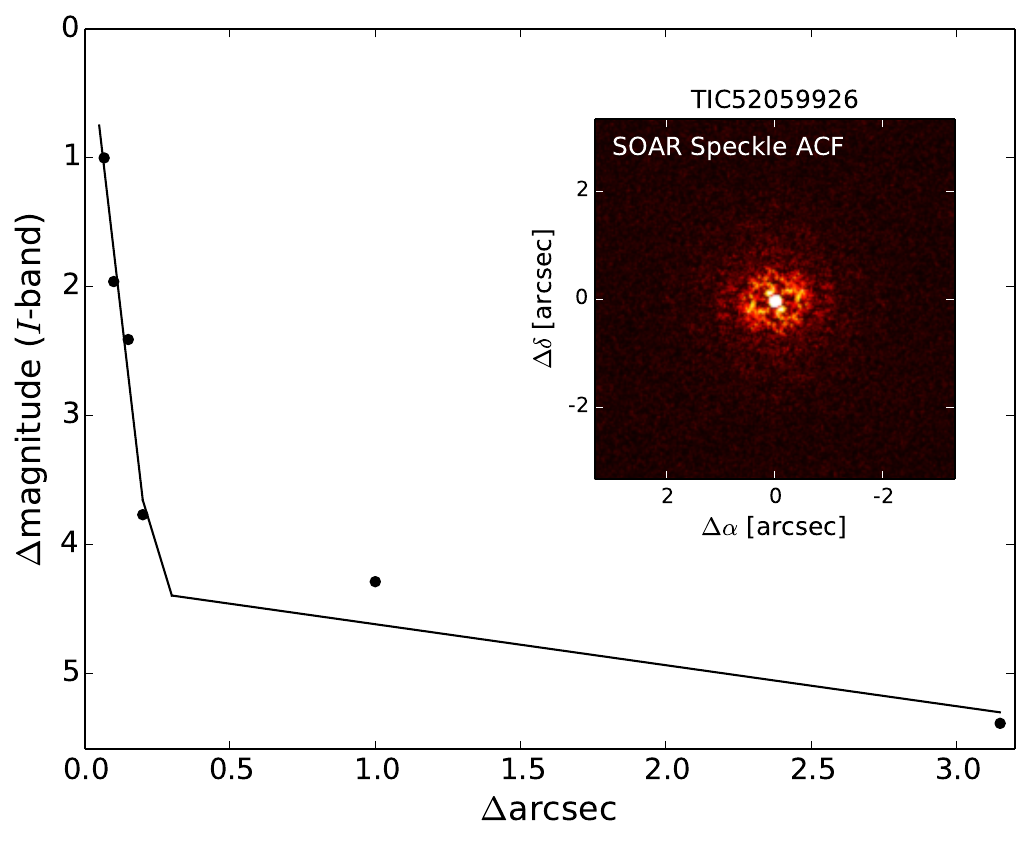}
\caption{SOAR contrast curve for Cousins $I$ band with a $6\arcsec \times 6\arcsec$ reconstructed image of the field.} \label{fig:speckle_image}
\end{figure}

On July 14, 2021, the High-Resolution Camera \citep[HRCam;][]{Tokovinin08} speckle interferometry instrument was employed on the Southern Astrophysical Research (SOAR) 4.1m telescope. The observation was made in the Cousins $I$ filter with a resolution of 36\,mas. The observation was carried out according to the observation strategy and data reduction procedures outlined in \cite{Tokovinin18} or \cite{Ziegler21}. The final reconstructed image achieves a contrast of ${\Delta}mag$=5.4 at a separation of $1\arcsec$, and the estimated PSF is 0.067\,arcsec wide. Fig. \ref{fig:speckle_image} plots a visual representation of the final reconstructed image. We found no companion with a contrast of ${\Delta}mag$ $\sim 4.5$ at projected angular separations ranging from 0.5 to $3.0\arcsec$ (55 to 332\,au).

\subsection{FEROS spectra}

We conducted a spectroscopic follow-up campaign with the FEROS spectrograph \citep{Kaufer1999} installed at the ESO-MPG 2.2 m telescope at La Silla Observatory. Observations were performed with the simultaneous calibration technique, in conjunction with a ThAr calibration lamp. FEROS data were reduced, extracted, and analyzed with the Ceres pipeline \citep{Brahm2017}, delivering RV and bisector span measurements. We summarize the spectroscopic observations in Table \ref{table:RV_s}.

\subsection{PLATOSpec spectra}

We conducted a spectroscopic follow-up campaign with the PLATOSpec spectrograph \citep{Kabath2025} installed at the ESO 1.52 m telescope at La Silla Observatory. PLATOSpec is a white pupil echelle spectrograph with spectral resolving power of $R\,=\,70,000$, covering the range of wavelengths from 380 to 700 nm. Observations were performed with the simultaneous calibration technique, in conjunction with a ThAr calibration lamp. PLATOSpec data were reduced, extracted, and analyzed with the Ceres pipeline \citep{Brahm2017}, delivering RV and bisector span measurements. We summarize the spectroscopic observations in Table \ref{table:RV_s}.

\begin{table}
\caption{Summary of Radial Velocity Measurements}             
\label{table:RV_s}      
\centering                          
\begin{tabular}{c c c c}        
\hline\hline                 
Target & N$_{obs}$ & Median $\sigma_{RV}$ & Observations dates \\    
\hline                        
   TIC\,9344899 & 10 & 16 & 2022 March 07 \\
   & & & 2022 October 31 \\
   \hline
   TIC\,52059926 & 35 & 18 & 2020 December 04 \\
   & & & 2023 August 25 \\
    \hline
   TIC\,13344668 & 19 & 17 & 2020 March 02 \\
   & & & 2022 April 15 \\
    \hline
   TIC\,63921468 & 25 & 23 & 2025 February 04 \\
   & & & 2025 September 14 \\
    \hline
    
\hline                                   
\hline
\end{tabular}
\end{table}

%
%

\section{Analysis} \label{sec:analysis}

\subsection{Modeling Stellar Parameters with iSpec and ARIADNE} \label{st_par}

We achieved a high signal-to-noise ratio by co-adding all high-resolution FEROS spectra after correcting for radial velocity shifts. This combined spectrum was utilized to derive the stellar parameters of our sample using the Spectroscopy Made Easy (SME) radiative transfer code \citep[{\tt SME};][]{Valenti96,Piskunov17}, integrated into the iSpec framework \citep{Blanco14,Blanco19}. Additionally, we modeled the spectra using MARCS atmospheric models \citep{Gustafsson08} and employed version 5 of the GES atomic line list \citep{Heiter15}. The iSpec software generates synthetic spectra and compares them with the observed spectrum, minimizing the chi-squared statistic through a nonlinear least-squares fitting algorithm (Levenberg-Marquardt) \citep{Markwardt09}.

For our analysis, we focused on the spectral region from 490 to 580 nm to ascertain the effective temperature ($T_{\rm eff}$), metallicity ($\rm [Fe/H]$), and projected stellar equatorial velocity ($v\sin{i}$). We applied the Bayesian parameter estimation code {\tt PARAM 1.5} \citep{DaSilva06,Rodrigues14,Rodrigues17} to derive the surface gravity ($\log{g}$) from PARSEC isochrones \citep{Bressan12}, incorporating the spectral parameters obtained and available photometric data (see Table \ref{tab1}), along with {\it Gaia} DR3 parallax measurements. This iterative procedure was executed until convergence to the final parameter values was achieved, which are summarized in Table \ref{table:stellar_par}.

\begin{figure*}
\centering
\sidecaption
\includegraphics[width=0.65\textwidth, trim= {0.0cm 0.0cm 0.0cm 0.0cm}]{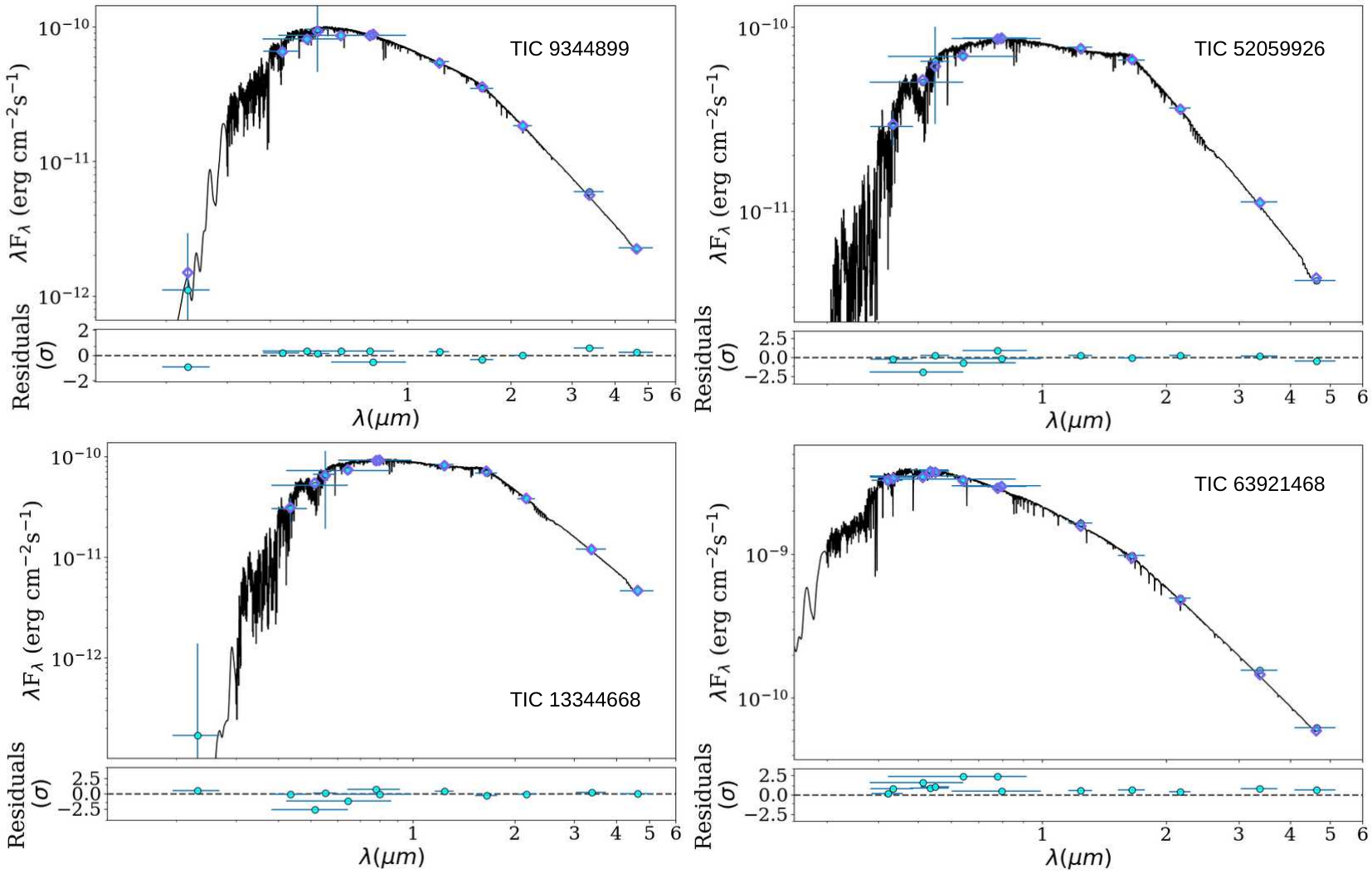}
\caption{SED fits for the systems in our sample. The blue points with vertical error bars show the observed catalog fluxes and uncertainties from {\it Gaia}, TESS, Tycho-2, 2MASS, and WISE, while the horizontal error bars illustrate the width of the photometric band. For illustrative purposes, we overplot the best-matching Phoenix atmospheric model \citep{Husser2013}.} \label{fig:SED}
\end{figure*}

To independently verify our derived stellar parameters, we modeled the spectral energy distribution (SED) with the Bayesian model averaging fitter, spectrAl eneRgy dIstribution bAyesian moDel averagiNg fittEr \citep[{\tt ARIADNE};][]{Vines22}. We determined $T_{\rm eff}$, $\log{g}$, metallicity ($\rm [Fe/H]$), stellar luminosity, and radius by combining 3D dust Bayestar maps for interstellar extinction \citep{Green19} with two atmospheric models: Kurucz \citep{Kurucz93}, and Ck04 \citep{Castelli2004}. Our priors were informed by the results obtained from iSpec. The final derived parameters, presented in Table \ref{table:stellar_par}, align well with previously established values. The SEDs are illustrated in Fig.~\ref{fig:SED}, featuring the best-fitting Phoenix model \citep{Husser2013}.

\begin{figure}
\centering
\includegraphics[width=0.45\textwidth, trim= {0.0cm 0.0cm 0.0cm 0.0cm}]{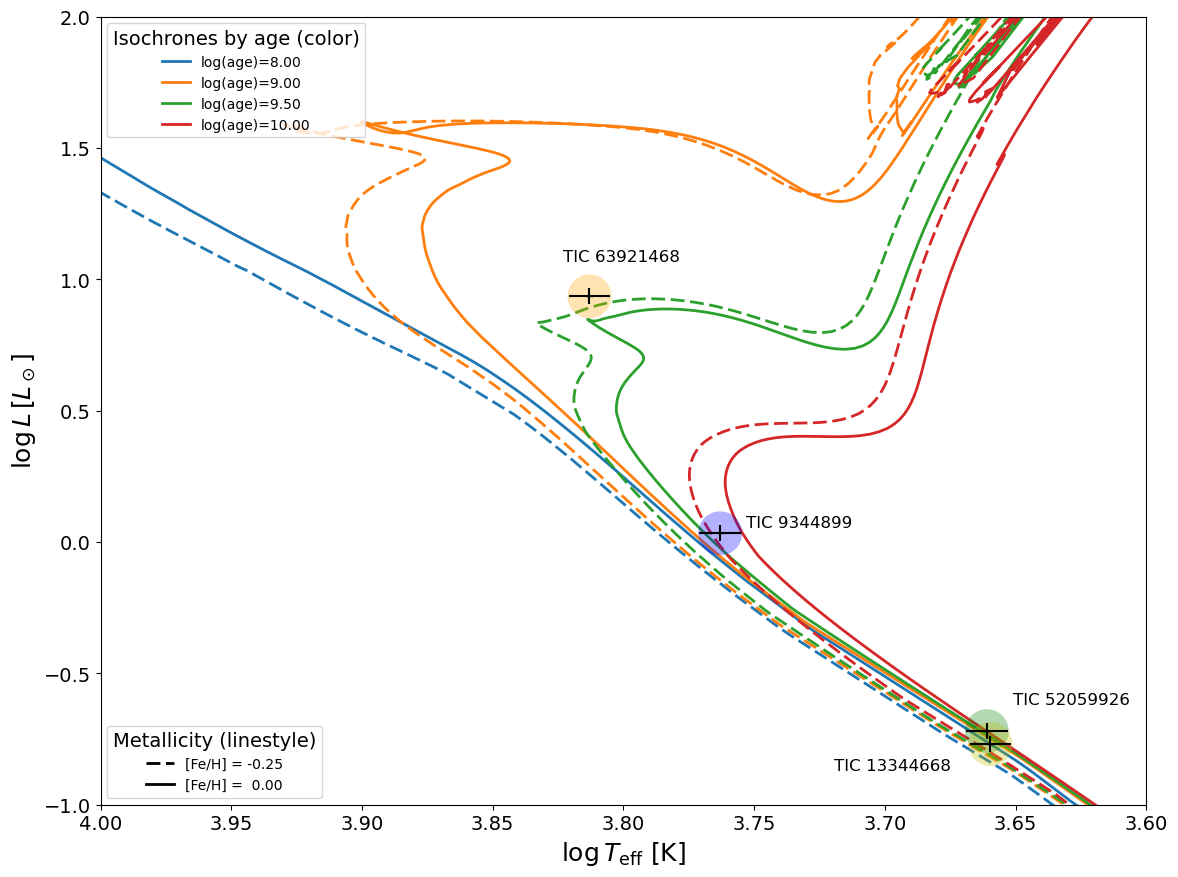}
\caption{Luminosity vs. effective temperature for the stars in our sample. Curves represent MIST isochrones for ages: 0.1\,Gyr (blue), 1.0\,Gyr (orange), 3\, Gyr (green), and  10.0\,Gyr (yellow). Different line types represent different metallicities. The red point represents the positions of stars with their error bars.} \label{fig:isochrones}
\end{figure}

To estimate the mass, radius, and age of the stars, we utilized MIST stellar evolutionary tracks \citep{Choi16} based on our stellar parameters and the available photometry. In Fig. \ref{fig:isochrones}, we overlay the luminosity and effective temperature of our stars with the MIST tracks. Spectral types were estimated using the latest version\footnote{\url{https://www.pas.rochester.edu/~emamajek/EEM_dwarf_UBVIJHK_colors_Teff.txt}.} of the empirical spectral type-color sequence as proposed by \cite{Pecaut13}.

\begin{table*}
 \centering
 \caption[]{Stellar parameters of TOI-2458.
 }
 \label{table:stellar_par}
\begin{adjustbox}{width=0.95\textwidth} 

	\begin{tabular}{lccccccr} 
		\hline
		\hline
& TIC\,9344899 & TIC\,52059926 & TIC\,13344668 & TIC\,63921468 \\
\hline
$\rm T_{eff}$ (K) & $5800 \pm 70$ & $4505 \pm 70$ & $4491 \pm 70$ & $6500 \pm 70$ \\
$[{\rm Fe/H}]$ (dex) & $0.07 \pm 0.05$ & $-0.08 \pm 0.05$ & $-0.20 \pm 0.05$ & $-0.22 \pm 0.05$ \\
$\log{g}$ (cgs) & $4.39 \pm 0.04$ & $4.62 \pm 0.02$ & $4.63 \pm 0.02$ & $3.85 \pm 0.02$\\
$v_{\rm rot} \sin{i_\star} $ (km/s) & $2.0 \pm 1.5$ & $1.5 \pm 1.5$ & $1.5 \pm 1.5$ & $11.0 \pm 1.5$\\
$M_\star$ ($\rm \mst$) & $0.98 \pm 0.08$ & $0.73 \pm 0.04$ & $0.70 \pm 0.03$ & $1.52 \pm 0.05$\\
$R_\star$ ($\rm \rst$) & $1.06 \pm 0.03$ & $0.72 \pm 0.02$ & $0.69 \pm 0.02$ & $2.35 \pm 0.05$\\
\hline
& ARIADNE analysis & \\
\hline
$\rm T_{eff} (K)$ & $5703 \pm 69$ & $4479 \pm 52$ & $4444 \pm 57$ & $6442 \pm 69$ \\
$[{\rm Fe/H}]$ & $0.10 \pm 0.10$ & $-0.07 \pm 0.10$ & $-0.18 \pm 0.10$ & $-0.22 \pm 0.10$ \\
$\log{g}$ & $4.38 \pm 0.04$ & $4.62 \pm 0.03$ & $4.63 \pm 0.03$ & $3.86 \pm 0.02$ \\
$L_\star$ (L$_{\odot}$) & $1.08 \pm 0.07$ & $0.19 \pm 0.01$ & $0.17 \pm 0.01$ & $8.61 \pm 0.43$ \\
\hline
& Other parameters & \\
\hline
Spectral type & G3V & K4.5V & K5V & F5V \\
Age (Gyr) & $6.1_{-2.9}^{+4.1}$ & $8.2_{-5.5}^{+5.2}$ & $2.7_{-2.2}^{+7.8}$ & $2.1_{-0.2}^{+0.2}$\\
\hline
\hline
	\end{tabular}
\end{adjustbox}
\smallskip\\
\end{table*}

\subsection{Frequency Analysis}\label{sec:Freq}

To differentiate the Doppler reflex motion from signals due to planetary candidates and stellar activity, we conducted a frequency analysis on the FEROS RVs and Bisector Inverse Slope (BIS) values derived from the Ceres pipeline. We computed Generalized Lomb-Scargle (GLS) periodograms \citep{Zechmeister09} for the time series data, establishing theoretical false alarm probabilities (FAP) at levels of 10\%, 1\%, and 0.1\%, as illustrated in Fig.~\ref{fig:per}. 

The periodogram revealed a significant signal for the TIC\,9344899 RVs, corresponding to periods of 8.6 days, aligning with the transiting planets identified through TESS photometry. Following the removal of this signal, the periodogram of the residual RVs exhibited no further significant signals, nor did the BIS periodogram reveal any noteworthy peaks.

In the case of the TIC\,52059926 and TIC\,13344668 systems, the planetary candidates remain undetermined in period due to TESS only capturing a single transit for each system. Analysis of the RV periodograms unveiled significant peaks at longer periods of 176 days and 142 days. Although additional significant peaks were also observed at even longer periods, the joint transit and RV modeling discussed in the subsequent section clarified the true period values, as highlighted in Fig.~\ref{fig:per}.

Notably, we detected a potential additional planet in the TIC\,52059926 system with a shorter period of 3.7 days, which is non-transiting. Nonetheless, further RV monitoring is essential to confirm this signal. This finding is particularly intriguing, as it could represent the first instance of a transiting brown dwarf having a less massive companion on the inner orbit.

\subsection{Joint RVs and transits modeling}\label{sec:RVs}
The modeling of both the transit light curves and radial velocity (RV) measurements was performed in a fully joint framework using the {\tt allesfitter} package \citep{Gunther21}. This software integrates several specialized tools: {\tt ellc} for transit and RV modeling \citep{Maxted16}, {\tt aflare} for stellar flare characterization \citep{Davenport14}, {\tt dynesty} for nested sampling \citep{Speagle20}, {\tt emcee} for Markov Chain Monte Carlo exploration \citep{Foreman13}, and {\tt celerite} for Gaussian process (GP) modeling of correlated noise \citep{Foreman17}. The flexibility of {\tt allesfitter} allows us to model a wide range of astrophysical effects, including multi-planet and multi-star systems, transit timing variations, phase curves, stellar activity, starspots, and flares, as well as a variety of systematic trends.

Before the fit, the light curves were corrected for stellar variability following the procedure described in Section~\ref{sec:TESS}. A quadratic limb-darkening law parameterized by $q_1$ and $q_2$ was adopted. The joint fit included our FEROS/PLATOSpec RV measurements, reduced according to the method described in Section~\ref{sec:observations}. The resulting posteriors are summarized in Table~\ref{tab:sec_fit}, while Table~\ref{tab:sec_der} lists the final transit and physical parameters, derived using the results from Table~\ref{tab:sec_fit}. The corresponding model fits are shown in Fig.~\ref{fig:transit_all}, Fig.~\ref{fig:transit}, and Fig.~\ref{fig:RVs}.

\begin{figure*}
\sidecaption
  \includegraphics[width=12cm]{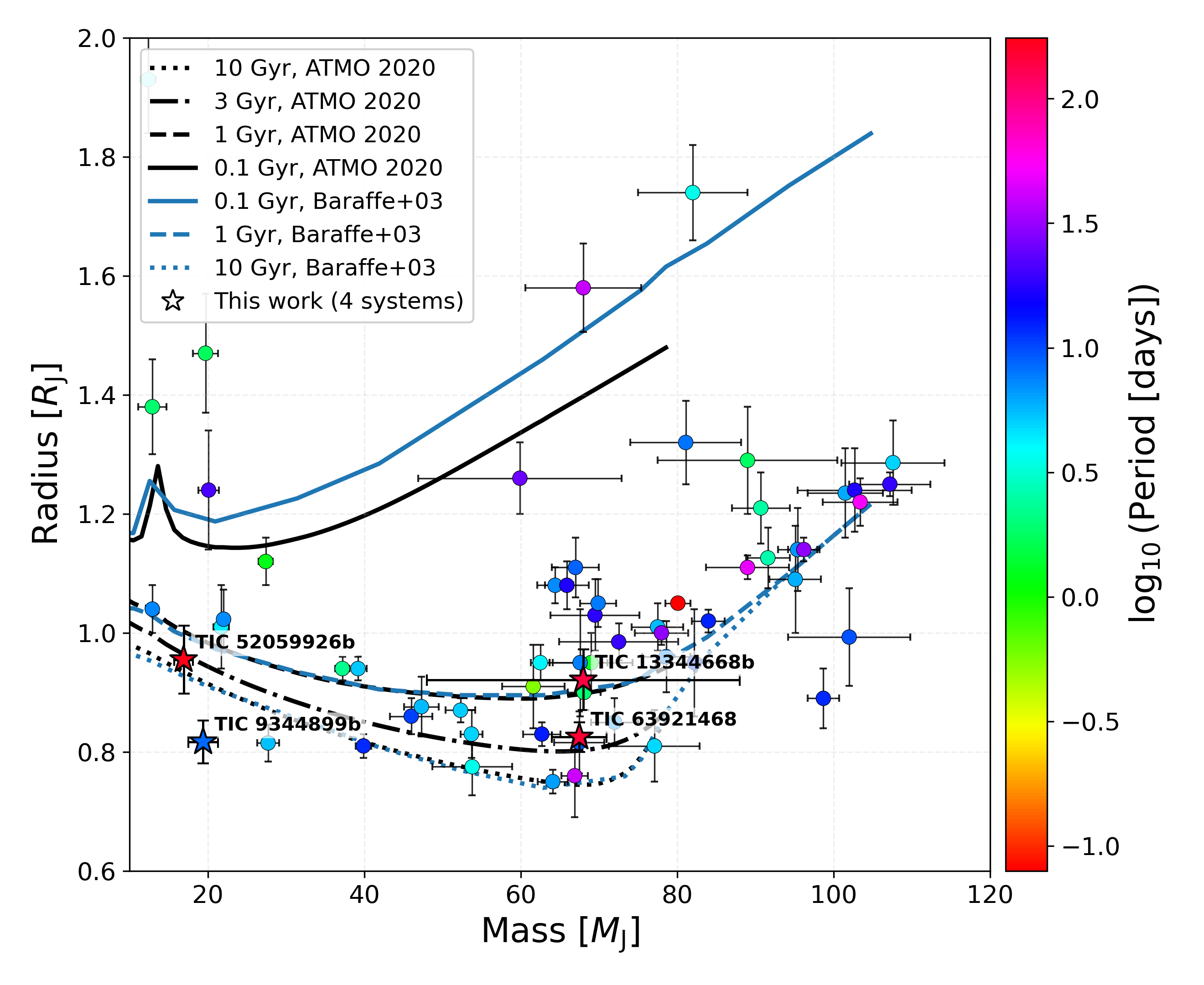}
     \caption{The population of known transiting brown dwarfs. We filtered out objects with large uncertainties in mass or radius. Star symbols highlight the positions of our systems. The plot is color-coded by orbital period. The mass-radius tracks from ... are utilized.}
     \label{fig:mr}
\end{figure*}

%
%

\section{Discussion} \label{sec:discussion}
\subsection{Mass-radius diagram}\label{tidals}

Combining data from the TESS mission with ground-based spectroscopy, we successfully confirmed the substellar nature of four candidates orbiting K, G, and F-type stars. Our findings indicate that two brown dwarfs (BDs) are situated near the lower mass boundary for BDs, while the other two exhibit masses approximately 70\,$M_{\rm J}$. Notably, TIC~13344668b is characterized by substantial mass uncertainties due to RV sampling that did not capture the minimum phase near 0.0, leading to multiple viable eccentricity solutions\footnote{We hope to cover this phase in the upcoming weeks during the referee phase.}. It is worth noting that TIC~63921468\,b has previously been identified as a transiting exoplanet candidate in the study conducted by \citet{Salinas25}.

Figure~\ref{fig:mr} illustrates the positions of our four companions (denoted by star symbols) within the brown-dwarf mass–radius diagram, overplotted against evolutionary tracks as presented by \citet{Baraffe2003} and the ATMO models \citep{Phillips2020} spanning 0.1, 1, 3, and 10~Gyr. For objects that are degenerate and undergoing cooling, the models indicate a weak dependence of radius on mass, ranging from approximately 0.8 to 1.1\,$R_{\rm J}$ across the mass interval of approximately 15 to 80\,$M_{\rm J}$. Furthermore, there is a monotonic contraction that occurs with age as internal entropy is gradually radiated. In contrast to hot Jupiters, the inflation driven by irradiation is anticipated to be minimal at brown dwarf masses. Consequently, deviations from the model envelopes typically signal the presence of unusual physical phenomena (such as ongoing tidal interactions, compositional variations, or deuterium burning) or systematic errors in the derived stellar parameters.

\textit{TIC~9344899\,b}, \textit{TIC~52059926\,b}, \textit{TIC~13344668\,b}, and
\textit{TIC~63921468\,b} fall within the 1--10~Gyr model envelopes, following the
canonical BD cooling sequence: at their measured masses, they cluster near
$\sim$0.8--1.0\,$R_{\rm J}$ with no compelling evidence for inflation. Their
periods (color scale in Fig.~\ref{fig:mr}) span from short to long; the lack of
radius excess at high equilibrium temperatures (TIC~9344899\,b) in this subset reinforces the view that stellar irradiation alone rarely produces large radius anomalies in the BD regime.

\textit{TIC~9344899\,b} lies on the compact edge of the brown-dwarf mass–radius relation (Fig.~\ref{fig:mr}). At $M\approx19\,M_{\rm J}$ its measured radius ($R\approx0.82\,R
_{\rm J}$) sits below the solar-composition ATMO/Baraffe evolutionary tracks at 10,Gyr, which generally approach a slow contraction asymptote of $\sim$\,0.9–1.0\,$R_{\rm J}$ in this mass regime. This places \textit{TIC~9344899\,b} in mild tension with standard isolated–cooling models for solar composition \citep[e.g.,][]{Baraffe2003,Phillips2020}. It is important to note that an underestimated radius may result from an imprecise dilution estimate, particularly given the crowded field surrounding the star. In our methodology (see Sec.~\ref{sec:TESS}), we employed the most reliable approach currently available, which has recently been utilized to correct the radii of numerous planets discovered by TESS \citep{Han2025}. The numerous ground-based photometry unfortunately cannot be used to confirm the transit depth due to insufficient coverage of the out-of-transit baseline.

A useful analogue is the recently validated transiting brown dwarf TOI-5422\,b ($27.7_{-1.1}^{+1.4},M_{\rm J}$, $0.815_{-0.021}^{+0.036},R_{\rm J}$, $P=5.38$\,d), whose discovery team explicitly describe it as slightly “underluminous” with respect to substellar evolution models at an old age ($8.2\pm2.4$\,Gyr) \citep{Zhang2025}. TOI-5422\,b demonstrates that short-period, old brown dwarfs with radii near $\sim$\,0.8\,$R_{\rm J}$ do exist; \textit{TIC~9344899\,b} would extend this compact tail to \emph{lower} mass, where most grids predict somewhat larger radii.

Physically, a slightly sub-model radius at fixed mass can be produced by (i) lower atmospheric/metallicity opacities that hasten cooling, or (ii) higher mean molecular weight (e.g., interior metal enrichment), each shrinking the radius by a few percent at Gyr ages. “Cold–start’’ initial entropies can also push toward smaller radii at early times, but for high-mass objects ($\geq5-10\,M_{\rm J}$) the hot–cold differences decay within \(\lesssim 10^8\) yr and are negligible by Gyr ages \citep{Spiegel2012}. The models used are based on standard, solar-composition, isolated-cooling grids that utilize a singular H/He equation of state along with specific opacity and cloud prescriptions. These foundational assumptions do not account for several factors that may introduce biases to the radii estimates at the few to ten percent level. Notable omissions include metallicity and cloud-dependent opacities, gradients in interior composition, and systematic discrepancies in the equation of state. Consequently, the actual lower envelope of radii may be somewhat more compact than the baseline predictions suggest. In particular, recent advancements in the dense H/He equation of state indicate slightly increased compressibility when compared to older datasets. This adjustment results in reduced adiabatic temperatures, accelerated cooling rates, and subsequently modestly smaller radii at fixed mass and age \citep{Chabrier2023}. The ongoing integration of such refined physical models, together with improved opacity and cloud treatment as well as inclusion of composition effects, is anticipated to enhance the precision of the theoretical lower envelope and may alleviate the discrepancies observed in compact systems such as TIC 9344899\,b and TOI-5422\,b.

\begin{figure}
\centering
  \includegraphics[width=0.45\textwidth, trim= {0.0cm 0.0cm 0.0cm 0.0cm}]{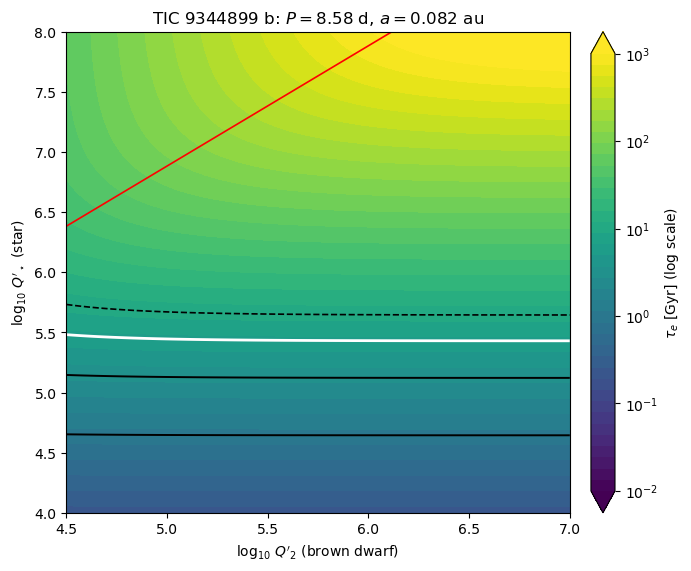}
     \caption{Circularization timescale \(\tau_e\) (Gyr, logarithmic color scale) as a function of the modified tidal quality factors of the brown dwarf and the star for TIC\,9344899. 
Black contours mark \(\tau_e=1,3\) Gyr (solid) and \(10\) Gyr (dashed); the white contour indicates the system median age; the red curve shows the boundary where tides in the companion and in the star contribute equally (\(\tau_{e,2}=\tau_{e,\star}\)). 
Contours are nearly horizontal in both panels, indicating that \(\tau_e\) is chiefly set by stellar dissipation over plausible \(Q'_2\).}
     \label{fig:tides}
\end{figure}

\subsection{Tidal interaction}\label{tidals}

Following \citet{Jackson2008}, in the small–eccentricity limit the eccentricity
damps approximately exponentially, $\dot e \simeq -e/\tau_e$, with
\begin{equation}
\frac{1}{\tau_e}
=\;\frac{63}{4}\,n\,\frac{M_\star}{M_2}\left(\frac{R_2}{a}\right)^{\!5}\frac{1}{Q'_2}
\;+\;\frac{171}{16}\,n\,\frac{M_2}{M_\star}\left(\frac{R_\star}{a}\right)^{\!5}\frac{1}{Q'_\star},
\label{eq:tau_e_jackson_n}
\end{equation}
where $n=\sqrt{G(M_\star+M_2)/a^3}$ is the mean motion, $Q'_2$ and $Q'_\star$
are the modified tidal quality factors of the companion and the star
($Q'\equiv 3Q/2k_2$, with $k_2$ the Love number), $M_\star$ ($M_2$) and
$R_\star$ ($R_2$) are the stellar (companion) mass and radius, and $a$ is the
semi-major axis.

It is often convenient to write the two contributions as separate timescales,
\begin{equation}
\tau_{e,2}=\frac{4}{63}\,Q'_2\left(\frac{M_2}{M_\star}\right)\left(\frac{a}{R_2}\right)^{\!5}\frac{1}{n},
\qquad
\tau_{e,\star}=\frac{16}{171}\,Q'_\star\left(\frac{M_\star}{M_2}\right)\left(\frac{a}{R_\star}\right)^{\!5}\frac{1}{n},
\label{eq:tau_e_parts}
\end{equation}
so that $1/\tau_e=1/\tau_{e,2}+1/\tau_{e,\star}$.
These expressions assume the equilibrium tide with constant phase lag, zero
obliquities, and the small-$e$ limit; $\tau_e$ is the e-folding time, not the
time to reach $e=0$.

We evaluate the eccentricity–damping time \(\tau_e\) for our 8\,d system TIC\,9344899 over the plausible tidal quality factors (\(\log Q'_2\sim4.5\!-\!6.5\) for BDs and
\(\log Q'_\star\sim4\!-\!8\) for stars;
\citealt{Heller2010,Beatty2018,Jackson2008,Penev2012,CollierCameron2018,Vissapragada2022}).

TIC 9344899 system falls in a regime where \emph{stellar} tides dominate the eccentricity damping for most of the plausible $Q'$ ranges; the decisive levers are the sizes and masses of the bodies. 
\textit{TIC~9344899\,b} orbits a near-solar host star with a radius of \(R_\star = 1.06\,R_\odot\). It has a more compact companion with a radius of \(R_2 = 0.82\,R_{\rm J}\), which is also located relatively far away, resulting in slower tidal interactions. The older age of TIC 9344899\,b, estimated at \(6.1^{+4.1}_{-2.9}\) billion years, partly compensates for its weaker tidal effects. Read directly from the $\tau_e$ maps in Fig. \ref{fig:tides}, the stellar $Q'_\star$ threshold
where $\tau_e\simeq$\,age is therefore \emph{similar}: $\log_{10}Q'_\star\approx5.5$.

For solar-type stars, both population studies and theoretical models generally suggest a higher value of \(Q'_\star\) (indicating less dissipation) on the main sequence, which typically falls within the range of $Q'_\star \sim 10^{5.5\text{–}6.5}$ \citep[][and references therein]{Ogilvie2014}. This implies that the circularization of orbits likely occurs more slowly than the median stellar age, unless the star exhibits unusually high dissipation or the system's age is at the upper end of the uncertainty range.

Given these efficiencies, the system does not permit a sharp distinction between
gentle (disk/in–situ) and impulsive (scattering/Kozai) pathways: the present low eccentricity is consistent with either, once plausible $Q'_\star$ and ages are allowed.
Targeted RV follow-up to search for outer perturbers would help to break this degeneracy.

\subsection{Brown dwarf population and eccentricity distribution}\label{pop}

\begin{figure}
\centering
\includegraphics[width=0.45\textwidth, trim= {0.0cm 0.0cm 0.0cm 0.0cm}]{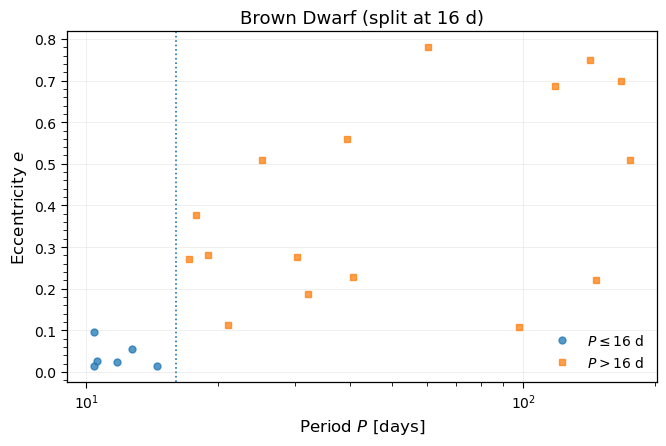}
\caption{Eccentricity vs. period distribution of transiting brown dwarfs with periods above ten days separated by the 16-d period boundary.} \label{fig:split}
\end{figure}

\begin{figure}
\centering
\includegraphics[width=0.45\textwidth, trim= {0.0cm 0.0cm 0.0cm 0.0cm}]{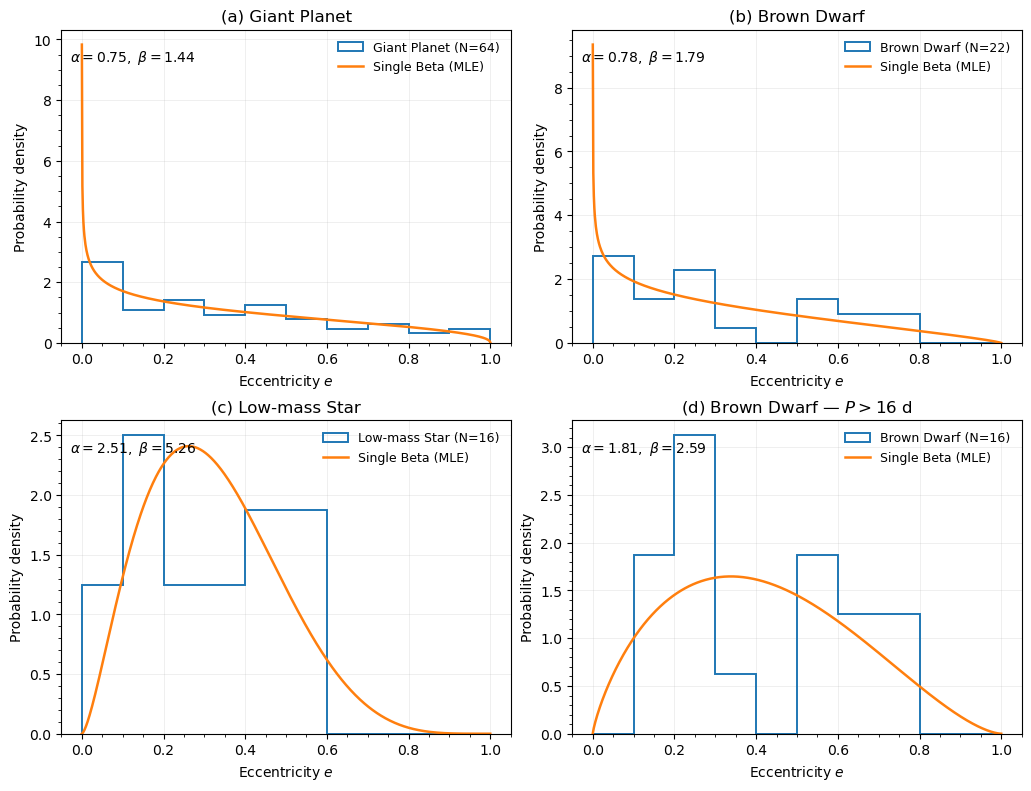}
\caption{Eccentricity distributions for (a) giant planets, (b) all transiting brown dwarfs ($P>10$\,d),
  (c) low-mass stellar companions, and (d) the BD subsample with $P>16$\,d.
  Step histograms use uniform bins in $e$; solid curves show the single-Beta maximum-likelihood fit
  to each panel (annotated by $(\alpha,\beta)$). Sample sizes are indicated in the legends.
  BDs ($P>16$\,d) resemble LMSs more than GTPs, with a higher, more concentrated mid-$e$ preference, whereas the all-BDs distribution is pulled toward the
  GTP-like shape by the short-period, low-$e$ objects. This visual comparison complements our statistical results (ES permutation $p=0.016$ for GTP vs.\ BD($>16$\,d); non-separation of BD($>16$\,d) from LMS at current $N$).} \label{fig:beta}
\end{figure}

Recent work on transiting long-period companions has argued that brown dwarfs (BDs) with $P>10$\,d exhibit an eccentricity distribution similar to that of warm/hot giant planets \citep{A_New_BD_Mstar_2025}. That conclusion rests on adopting $P>10$\,d as a practical boundary for ``negligible tides'' and then comparing the BD $e$ distribution to that of giant planets (GTPs) and low-mass stars (LMSs). In what follows, we reproduce the $P>10$\,d cut for comparability, but we also examine the period-eccentricity structure of the BD sample and find that the apparent GTP-like $e$ distribution is largely driven by a small subset of six systems at $10\!<\!P\!\le\!16$\,d with very low eccentricities (see Fig. \ref{fig:split} and Fig. \ref{fig:beta}).

Concretely, across four groups: GTPs, all BDs ($P\geq10$\,d), LMSs, and BDs with $P\geq16$\,d, we obtain (median $e$, 16–84\%): GTPs $(0.294,,0.035$–$0.670)$, all BDs $(0.250,,0.037$–$0.641)$, LMSs $(0.312,,0.160$–$0.512)$, BD($P\geq16$\,d) $(0.329,,0.200$–$0.693)$. Two-sample tests show that most pairs are not distinguishable at current $N$, but a shape-sensitive Epps–Singleton test flags differences between GTPs and LMSs and between GTPs and BDs ($P\geq16$\,d), while GTPs vs “all BDs” is not rejected. Parametric cross-checks with single-Beta fits illustrate the same shift (GTPs: $\alpha=0.75,\ \beta=1.44$; BDs ($P\geq16$\,d): $\alpha=1.81,\ \beta=2.59$), albeit with mixed information-criterion support given small samples (weak $\Delta$AIC\,$<0$ for separate fits; $\Delta$BIC\,$>0$ favoring a pooled model at current $N$). Taken together, the simplest takeaway is:

\begin{quote}
\emph{When short-period ($10$–$16$\,d) low-$e$ BDs are set aside, the $P\geq16$\,d BD population looks more LMS-like than GTP-like in eccentricity. Both the BD and LMS populations exhibit signs of bimodality; however, additional systems are needed for better statistical evaluation.}
\end{quote}

We asked whether the low-eccentricity BD systems at $P=10$--$16$\,d exhibit any shared host or companion properties that would mark them as a distinct formation channel. They do not: the host masses, radii, and metallicities span wide ranges, and the companions likewise cover a broad swath in $M_2$, and mass ratio $q$. In particular, [Fe/H] and $q$ show no significant offsets relative to longer-period BDs, and the only hint is a weak tendency toward lower $M_2$ and smaller $R_2$ at shorter periods (marginal at best). Lacking a clean demographic signature, the simplest explanation for their low eccentricities is \emph{partial tidal damping} at these separations.

A common back-of-the-envelope is the small-$e$ circularization time $\tau_e$ evaluated at the present semi-major axis (e.g., constant-phase-lag scaling; \citealt{Jackson2008}). Scanning wide dissipation grids $Q'_\star,Q'_2\in[10^4,10^8]$, we find that only two of six systems reach $\tau_e<10$,Gyr at the most dissipative edge ($Q'_\star\sim10^4$); for $Q'_\star\ge10^5$, all have $\tau_e>10$,Gyr. However, as expected, these small-$e$ estimates may understate early damping, because tidal rates increase steeply at high $e$.

We therefore integrated the standard constant-time-lag (CTL) equations for coupled star+BD tides with a pseudo-synchronous BD spin \citep{Hut1981,Jackson2008}. To set the initial orbit after an excitation phase, we adopted a constant-periastron start, which is a neutral, widely used way to connect high-$e$ excitation to subsequent tidal cooling; once excitation ceases, tides act at the existing periastron distance $r_p$.

For a modest $e_0=0.25$, we find for our 6 systems:

Fast/likely: TOI-1406b and TOI-4776b have broad ranges of $Q'\star$ in $[10^4,10^6]$ with $t_{\rm damp}(e_0\rightarrow e_{\rm obs})\lesssim5$\,Gyr.

Borderline: KOI-205b and TOI-5389Ab typically cool within $\sim5$\,Gyr near the more dissipative end $Q'_\star\leq10^{4.5-4.75}$.

Slow: LHS,6343b and TOI-1278b need $\gtrsim5$–$10$\,Gyr unless $Q'_\star$ is near $10^4$.

In our setup, the CTL damping rate gains a strong \emph{high-$e$} boost (through the Hut
eccentricity functions) but suffers an $a^{13/2}$ penalty from the inflated starting
semi-major axis. For \emph{modest} excitation ($e_0\!\sim\!0.25$; $a_0/a_{\rm now}=1/(1-e_0)=4/3$),
the high-$e$ boost typically \emph{outweighs} the $a^{13/2}$ penalty, so the
full CTL time can be \emph{shorter} than the small-$e$ $\tau_e$ evaluated at today’s $a$. By contrast, for \emph{strong} excitation
($e_0\!\gtrsim\!0.6$; $a_0/a_{\rm now}=2.5$), the $a^{13/2}\!\approx\!2.5^{6.5}$ and penalty dominates, making the CTL time \emph{longer} than the local small-$e$ estimate and pushing several systems beyond a Hubble time.

The fastest channels generally occur when stellar dissipation dominates (low $Q'\star$). Interestingly, our three companions with the weakest tidal interactions (TOI-5389\,Ab LHS\,6343\,b and TOI-1278\,b) are found around M-dwarf hosts. M-dwarfs being deeply convective, plausibly sit toward the more dissipative end \citep{Gallet2017}. Within these ranges, tides can plausibly explain the low $e$ for most—and potentially all—of the $10$–$16$\,d systems if their post-excitation eccentricities were modest ($e_0\sim0.2$–$0.3$) and $Q'\star$ is on the efficient side of empirically allowed values (especially for M dwarfs). Stronger initial excitation ($e_0\gtrsim0.6$) inflates $a_0$ (via $a_0\propto1/(1-e_0)$) and slows CTL damping roughly as $a^{13/2}$, pushing several systems beyond a Hubble time.

\smallskip
\noindent\textbf{Implication.} The GTP-like appearance of “BDs with $P>10$\,d” in prior work is largely set by the six low-$e$ objects at 10–16\,d. Those systems do not share a clean demographic signature but do sit where tides are most effective; full CTL integrations show that modest post-excitation $e_0$ and efficient (but empirically plausible) stellar dissipation suffice for all of them. Beyond 16\,d, transiting BDs resemble LMS binaries, and our three new $P!>!100$\,d systems strengthen the LMS-like high-$P$ tail.

\subsection{Host metalicity}\label{metalicity}

The planet--metallicity correlation is well established for close-in giant planets \citep[e.g.][]{Santos2004,Fischer2005,Petigura2018}, while stellar-mass companions show no or even negative preference for metal-rich hosts \citep[e.g.][]{Raghavan2010,Moe2019}. The goal of this section is to examine where brown dwarfs lie along this axis.

\begin{figure}
\centering
\includegraphics[width=0.45\textwidth, trim= {0.0cm 0.0cm 0.0cm 0.0cm}]{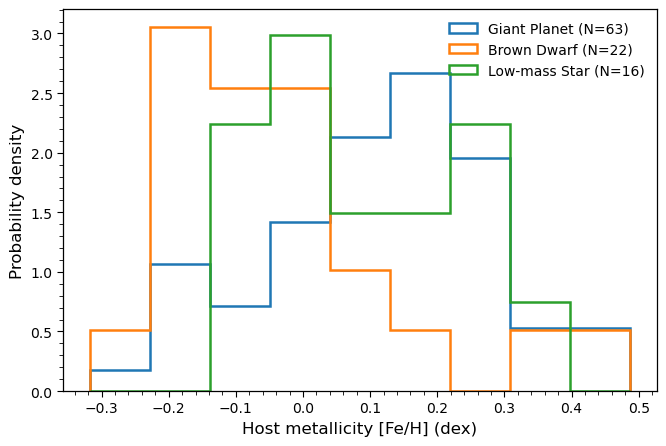}
\caption{Host [Fe/H] distributions for GTPs, LMSs, and BDs ($P>16$\,d).
BD hosts are more metal-poor than GTP hosts (multiple tests, Holm-adjusted
$p\le0.04$) and are not separable from LMS hosts at current $N$, consistent with the LMS-like behavior seen in eccentricities.} \label{fig:met}
\end{figure}

We compared the host metallicity distributions of three groups: GTPs, BDs, and LMSs. Median host metallicities are (see Fig. \ref{fig:met}):
\[
\text{GTPs}:\ +0.14~{\rm dex},\quad
\text{BDs}:\ -0.06~{\rm dex},\quad
\text{LMSs}:\ +0.06~{\rm dex},\quad
\]
with interquartile ranges of roughly $\simeq$0.20--0.25\,dex per group. Multiple two-sample tests reject equality for GTPs vs.\ BDs ($P{>}16$\,d) at the $p\!\sim\!10^{-2}$ level (KS $p_{\rm adj}=0.037$, CvM $p_{\rm adj}=0.007$, Epps--Singleton $p_{\rm adj}=0.004$, Anderson--Darling $p_{\rm adj}=0.009$). All other pairwise comparisons are consistent with equality at the current $N$. These metallicity patterns independently support our eccentricity-based conclusion that: \emph{long-period brown dwarfs resemble LMS companions more than giant planets}.

Population plots in \citet{Vowell2025} show that most \emph{short-period} transiting BDs---both below and above the often-quoted $40\,M_{\rm J}$ division---cluster around metal-rich hosts, following the same overall planet--metallicity trend as hot/warm Jupiters. By contrast, \emph{long-period} BD hosts skew metal-poor and become indistinguishable from low-mass stellar (LMS) companions within present uncertainties. We also do not see any metallicity bifurcation at $40\,M_{\rm J}$ within the current sample size: both lower- and higher-mass short-period BDs are abundant around metal-rich stars, whereas at long periods both mass ranges shift toward more field-like, typically lower metallicities. This pattern points to a \emph{period-coded} mixture rather than a mass-coded one.

Interestingly, an analogous period dependence is observed for the population of hot and cold Jupiters. \citet{Maldonado2018} used a large sample of stars and found that stars with hot Jupiters have higher metallicities than stars with cool, distant gas-giant planets, which they speculate as evidence for distinct formation channels for hot and cold Jupiters. In their scenario, planets in metal-poor discs either form farther out or form later and do not have time to migrate as far as planets in metal-rich systems. In the brown-dwarf regime, we intentionally avoid importing the same detailed conclusions, because such scenarios must be tuned separately for different seed mechanisms (core accretion for giant planets vs.\ fragmentation for BDs) and therefore add unnecessary complexity. By contrast, metallicity-dependent \emph{inward delivery and survival}---via more massive and longer-lived viscous discs in metal-rich systems---operates regardless of the initial formation channel. Metal-rich systems have longer-lived, more massive/viscous discs \citep{Yasui2009,Ercolano2010}, enabling disc-phase delivery and damping that yield a higher survival fraction of massive companions at small radii, whereas metal-poor systems disperse discs earlier, pushing inward delivery (if it occurs at all) into high-eccentricity, low-survival channels \citep[e.g. reviews and syntheses in][]{Dawson2018,Fortney2021}. This metallicity-dependent delivery–survival pipeline naturally explains the metal-rich excess among short-period companions without requiring a specific brown-dwarf formation channel.

Long-period BDs, on the other hand, look more like a stellar-binary–style channel (core/disc fragmentation) in which host metallicity only weakly influences the companion rate. Population studies of binaries show that close-binary fractions do not increase with [Fe/H] in the way giant-planet frequencies do and can even anticorrelate \citep{Moe2019}. Consistent with this, the eccentricity distribution of \emph{long-period} ($10\!\lesssim\!P\!\lesssim\!10^3$~d) transiting BDs is not planet-like: instead of peaking at $e\simeq0$ as warm Jupiters do, it is potentially bimodal, with one peak near $e\simeq0.2$ and a second near $e\simeq0.7$, quite resembling the distribution of close stellar binaries. Some fraction of the modest-eccentricity component may already be imprinted during disc dispersal: secular interactions between a companion inside an inner cavity and a massive, mildly eccentric outer disc can resonantly excite eccentricities even if the disc itself is only weakly eccentric \citep[e.g.][]{Li2023}. However, it is unlikely to be the sole origin of all eccentric systems. We therefore interpret the observed distribution more broadly as the superposition of (i) dynamically quieter systems—objects that migrated and circularised quiescently in or near the disc plane and/or experienced only mild secular evolution, yielding low to moderate eccentricities—and (ii) systems that were later excited to high eccentricities by binary-like secular processes (Kozai–Lidov cycles, scattering, or secular chaos in higher-order multiples) once the disc had dispersed. In this view, the short-period metallicity break is set primarily during the disc phase, by which companions are delivered to, and stabilised at, very short periods in metal-rich discs, whereas the bimodal eccentricity structure at longer periods mainly records subsequent, largely metallicity-insensitive binary-style secular evolution.

Spin–orbit measurements of brown-dwarf systems provide a complementary constraint on the migration geometry. All transiting BDs with Rossiter–McLaughlin measurements are consistent with low projected obliquities, even when their orbits are substantially eccentric: TOI-2119~b ($P\!=\!7.2$~d, $e\!\simeq\!0.3$), TOI-2533~b ($P\!\approx\!21$~d, $e\!\approx\!0.25$), and HIP~33609~b ($P\!=\!39$~d, $e\!=\!0.56$) all exhibit aligned or nearly aligned orbits within current uncertainties \citep{Ferreira2024,Doyle2025,Vowell2025}. This combination of a broad, binary-like eccentricity distribution and a narrow obliquity distribution points to eccentricity excitation that is predominantly \emph{coplanar} with the natal disc plane—such as secular interactions and scattering in hierarchically structured, but only mildly inclined, multiples. In this picture, most brown dwarfs inherit a disc-aligned configuration from their fragmentation and disc-interaction phase, with subsequent binary-style secular evolution broadening the eccentricity distribution without strongly tilting the orbital plane.

Short-period transiting brown dwarfs are therefore most naturally explained by inward delivery rather than in\mbox{-}situ formation at few-day orbits. Gravitational instability (GI) and disc fragmentation (stellar-/disc-like formation) operate preferentially at larger distances (tens to hundreds of au), not at $\lesssim 0.1$~au where cooling is inefficient and irradiation is strong \citep{KratterLodato2016}. Close binaries and BD companions at few-day periods are thus expected to have formed wider and then hardened/migrated inward through some combination of gas torques and multi-body dynamics \citep{Tokovinin2020CloseBinaries}. Inward evolution need not proceed as a single, smooth Type-II track: hydrodynamic simulations show that migration, resonant capture, and planet–planet scattering can all operate while the gas disc is still present, and that interaction with the disc can circularise or stabilise surviving companions after the chaotic phase, reducing the incidence of ejections and leaving some companions on bound, moderately eccentric orbits that continue to migrate inward \citep[e.g.][]{Moeckel2008,Marzari2010}. In this sense, even “high-eccentricity migration” channels can be disc-assisted when they operate during the gas phase, and the survival of a short-period BD is already contingent on the disc having enough mass and lifetime to both deliver and stabilise the companion.

The metallicity dependence then enters most directly through the disc properties rather than through the subsequent dynamical heating. High-eccentricity migration channels that occur after disc dispersal (such as Kozai–Lidov cycles or secular chaos in stellar or BD multiples) play a role in shaping the overall brown dwarf desert. However, there is no compelling reason to expect these post-disc processes to be strongly influenced by metallicity. The relevant perturbers in these cases are typically other stars or substellar companions that formed through fragmentation; their occurrence rates do not increase with higher metallicity and may even anticorelate \citep{Moe2019}. Consequently, while these channels can effectively broaden the distributions of semi-major axes and eccentricities, they do not inherently create a strong metallicity–period boundary on their own.

By contrast, \emph{disc-driven migration and damping} of a high-mass companion tie short-period survival much more directly to the structure and lifetime of the protoplanetary gas disc, properties that are empirically and theoretically linked to metallicity, with low-metallicity environments showing systematically shorter disc lifetimes and lower gas masses \citep{Yasui2009,KleyNelson2012}. In this framework, the metal-rich excess of short-period BDs reflects the larger window, in high-metallicity discs, for any combination of disc-assisted migration and early dynamical excitation to bring brown dwarfs safely to few-day orbits.

A recent \textit{TESS} timing analysis of transiting brown dwarfs reports \emph{no statistically significant transit-timing variations} attributable to additional companions \citep{Wang2025_ExoEchoBDTTV}. Throughout, by TTVs we mean \emph{three-body} signals from nearby perturbers in or near resonance \citep{Agol2005,Lithwick2012}, not long-term tidal-decay drifts. This null detection is \emph{consistent} with inward, disc-driven delivery by a gap-opening, high-mass companion: such bodies push material away from their orbit and carve deep gaps, tending to deplete the immediate vicinity of long-lived, close neighbours where three-body TTVs would be most detectable \citep{KleyNelson2012}. For context, early population studies found that hot Jupiters rarely show nearby companions/TTVs \citep{Steffen2012}, with the few companion-rich systems likely tracing in-situ or very gentle migration channels that preserve neighbours \citep{Becker2015,Almenara2016,Canas2019_Kepler730,Subjak2025_TOI2458}. In our framework, \emph{short-period} BDs near the cavity should likewise rarely show close three-body TTVs, as found by \citet{Wang2025_ExoEchoBDTTV}, whereas for \emph{long-period} transiting BDs the presence of both inner and outer gas during the nebular phase allows convergent migration and resonant neighbours in principle; such companions are, however, most plausibly relatively massive joint-gap neighbours, while low-mass near-resonant planets remain disfavoured by the BD’s wide gap and large chaotic zone.

Altogether, the most straightforward conclusion is that metal-rich systems are over-efficient at disc-delivering and retaining massive companions close in (hence the metal-rich short-period tail), whereas metal-poor systems preferentially populate the long-period BD regime that follow-up is now uncovering.

Finally, we point out several limitations in the current study. Metallicities come from heterogeneous spectroscopic pipelines and carry systematics, and small $N$ makes many pairwise tests non-rejectable. The case would be strengthened by (i) a substantially larger long-period transiting BD and LMS sample, and (ii) homogenising the host stellar parameters used in future population-level analyses.

\section{Summary} \label{sec:summary}

We report the discovery of four new transiting brown dwarfs, three of which exhibit orbital periods exceeding 100 days. Prior to this study, only two such long-period transiting brown dwarfs were known; thus, our findings effectively double the existing sample and enhance our understanding of the high-period tail of the population. Notably, all three long-period objects are found to orbit stars with subsolar metallicity, aligning with the broader trend observed for long-period brown dwarfs. 

In our analysis, we utilized catalogs of giant planets, brown dwarfs, and low-mass stellar companions with periods greater than 10 days to compare their eccentricity distributions. The inclusive dataset for all brown dwarfs with periods exceeding 10 days initially reflects the giant planet-like eccentricity distribution; however, this is influenced by six nearly circular objects within the 10 to 16-day period range. Upon excluding these objects, the sample of brown dwarfs with periods exceeding 16 days demonstrates a shift towards the regime of low-mass stars.

We investigated whether the 10–16 d brown dwarfs exhibit distinct demographic characteristics. Our analysis indicates no significant differences in host metallicity [Fe/H], mass ratio \( q \), or radii, suggesting that tidal interactions, rather than a unique formation process, are at play. The small-eccentricity circularization times calculated at their current semimajor axes appear prolonged. However, comprehensive circularization time law (CTL) integrations initiated from a moderate eccentricity (\( e_0 \sim 0.25 \)) with a conservative constant-periastron initialization demonstrate that all six systems have the potential to achieve their observed eccentricities within approximately 10 billion years. Notably, this process may necessitate lower values of \( Q'_\star \) for certain systems. Therefore, it is plausible that mild tidal interactions contribute to the low eccentricities observed among these short-period brown dwarfs, thereby influencing the distribution of eccentricities.

Host metallicities support the picture. Long-period brown dwarfs are predominantly found orbiting stars that exhibit significantly lower metallicity compared to long-period giant planet hosts. This observation aligns statistically with the characteristics of low-mass star hosts. In contrast, short-period brown dwarfs—regardless of whether they are below or above the frequently cited $40,M_{\rm J}$ threshold—are concentrated around metal-rich stars, reflecting the metallicity distribution observed in hot Jupiters. The lack of a distinct [Fe/H] bifurcation at $40,M_{\rm J}$ argues against a purely mass-coded split and instead favors a period-coded mixture. This conclusion is further substantiated by the analogous metallicity trends noted between hot and cold Jupiters. Metal-rich disks (more massive/viscous, longer-lived) preferentially deliver and retain massive companions at small radii, independent of whether the seeds formed via core accretion or fragmentation.

In summary, our identification of three new brown dwarfs with periods greater than 100 days highlights the expansion of the metal-poor, long-period distribution, where these brown dwarfs exhibit characteristics akin to low-mass stellar binaries in terms of both the eccentricity distribution $p(e)$ and [Fe/H]. Conversely, the short-period, metal-rich distribution likely reflects the influence of metallicity on inward migration and retention. Expanding the sample of transiting brown dwarfs and low-mass stars and homogenizing host metallicities will be critical for evaluating single- versus two-component models of $p(e)$ and for clarifying the relative contributions of formation and migration processes in shaping the inner and outer boundaries of the brown dwarf desert.

%
%
\begin{acknowledgements}
J.\v{S}. would like to acknowledge the support from GACR grant 23-06384O.
A.J. acknowledges support from Fondecyt project 1251439.
This work was supported by ANID Fondecyt n. 1251299 and Basal CATA FB210003.

Funding for the TESS mission is provided by NASA's Science Mission Directorate.

We acknowledge the use of public TESS data from pipelines at the TESS Science Office and at the TESS Science Processing Operations Center.

Resources supporting this work were provided by the NASA High-End Computing (HEC) Program through the NASA Advanced Supercomputing (NAS) Division at Ames Research Center for the production of the SPOC data products.


This paper includes data collected by the TESS mission that are publicly available from the Mikulski Archive for Space Telescopes (MAST).


This research has made use of the Exoplanet Follow-up Observation Program (ExoFOP; DOI: 10.26134/ExoFOP5) website, which is operated by the California Institute of Technology, under contract with the National Aeronautics and Space Administration under the Exoplanet Exploration Program.


This work makes use of observations from the LCOGT network. Part of the LCOGT telescope time was granted by NOIRLab through the Mid-Scale Innovations Program (MSIP). MSIP is funded by NSF.

This work is based in part on data collected under the NGTS project at the ESO La Silla Paranal Observatory. The NGTS facility is operated by a consortium institutes with support from the UK Science and Technology Facilities Council (STFC) under projects ST/M001962/1, ST/S002642/1 and ST/W003163/1.

This research has made use of the Simbad and Vizier databases, operated at the centre de Donn\'ees Astronomiques de Strasbourg (CDS), and of NASA's Astrophysics Data System Bibliographic Services (ADS).

This work has made use of data from the European Space Agency (ESA) mission {\it Gaia} (\url{https://www.cosmos.esa.int/gaia}), processed by the {\it Gaia} Data Processing and Analysis Consortium (DPAC, \url{https://www.cosmos.esa.int/web/gaia/dpac/consortium}). Funding for the DPAC
has been provided by national institutions, in particular, the institutions participating in the {\it Gaia} Multilateral Agreement.

This work made use of \texttt{TESS-cont} (\url{https://github.com/castro-gzlz/TESS-cont}), which also made use of \texttt{tpfplotter} \citep{Aller20} and \texttt{TESS-PRF} \citep{Bell22}.
\end{acknowledgements}

\bibliographystyle{aa}
\bibliography{astro_citations}

\onecolumn
\begin{appendix}
\section{Additional material}

\begin{figure*}[!htbp]
\centering
\sidecaption
\includegraphics[width=0.58\textwidth, trim= {0.0cm 0.0cm 0.0cm 0.0cm}]{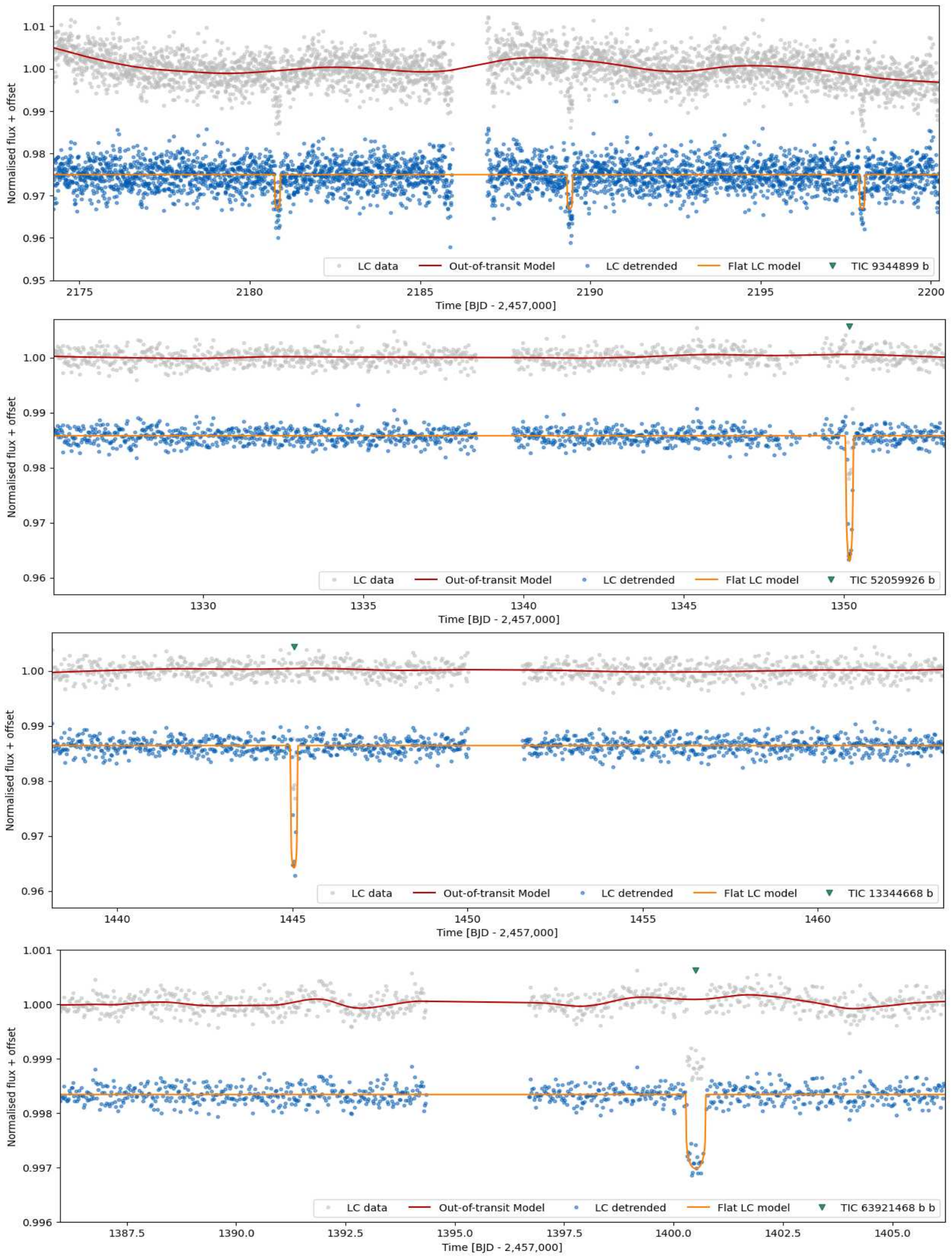}
\caption{TESS LCs for stars in our sample downloaded with the {\tt lightkurve} tool. Gray points represent the TESS observations, while red lines correspond to the out-of-transit GP models created with {\tt citlalicue} to capture the variability in the LC. Datasets were divided by these models, leading to a flattened TESS LC (blue points) with the transit model (orange line). Green triangles indicate the positions of transits. } \label{fig:tess_lc}
\end{figure*}

\begin{figure*}
\centering
\includegraphics[width=0.79\textwidth, trim= {0.0cm 0.0cm 0.0cm 0.0cm}]{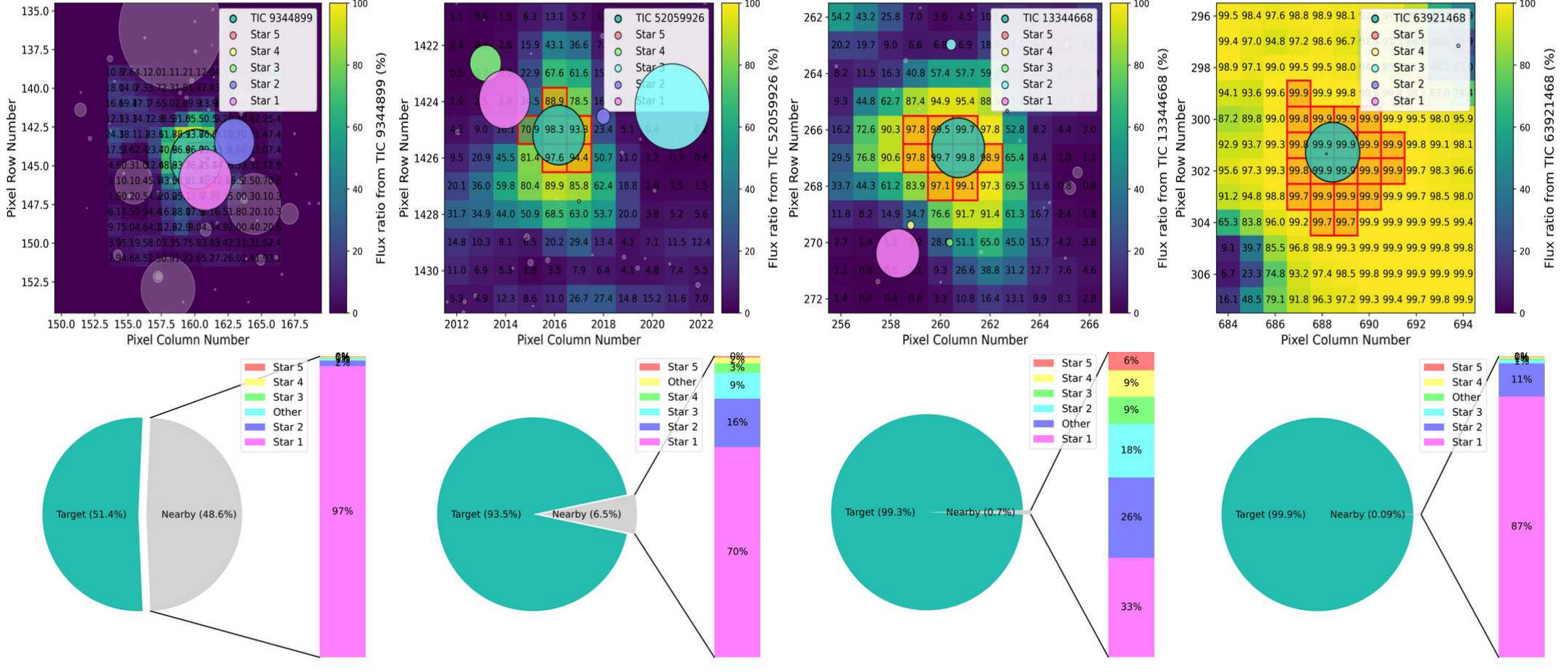}
\caption{Top: Heatmap illustrating the percentage of the flux from stars in our sample that falls within each pixel of the TESS TPF image using {\it Gaia} DR2/DR3 catalogs. Bottom: Pie chart representing the flux from the target and nearby stars inside the photometric aperture.} \label{fig:field_image}
\end{figure*}

\begin{figure*}
\sidecaption
  \includegraphics[width=12cm]{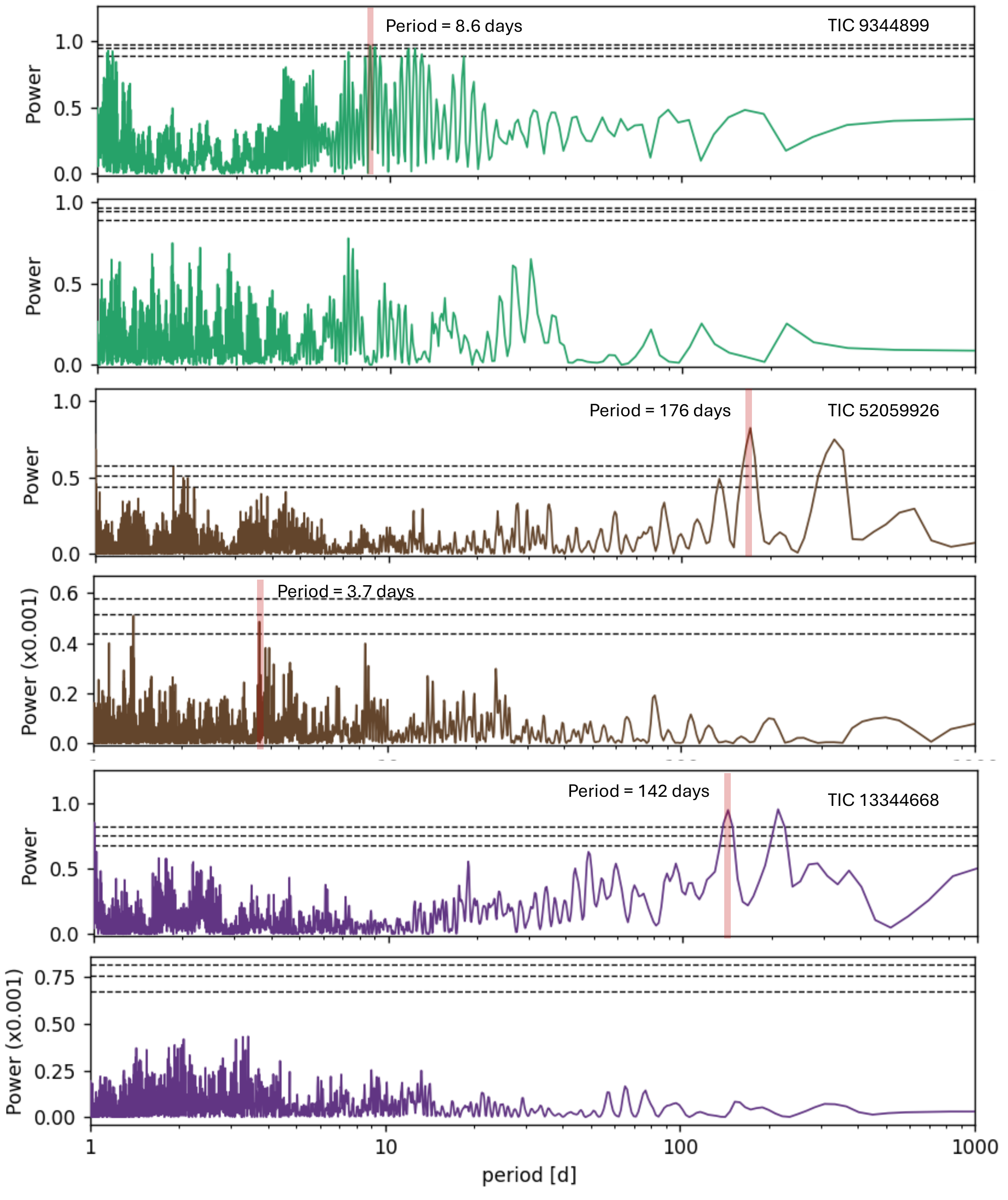}
     \caption{Generalized Lomb-Scargle periodograms of FEROS RVs of stars in our sample from top to bottom: (a) TIC\,9344899 RVs, (b) TIC\,9344899 RVs minus transiting planet signal, (c) TIC\,52059926 RVs, (d) TIC\,52059926 RVs minus planet signal, (e) TIC\,13344668 RVs, (f) TIC\,13344668 RVs minus planet signal. The vertical red lines highlight the orbital period of the companions. Horizontal dashed lines show the theoretical FAP levels of 10\%, 1\%, and 0.1\% for each panel.}
     \label{fig:per}
\end{figure*}

\begin{table*}
 \centering
 \caption{Fitted parameters from the {\tt allesfitter} analysis.} 
 \label{tab:sec_fit}
 \begin{tabular}{lcccccc}
    \hline
    \hline
    Parameter & TIC 9344899 & TIC 52059926 & TIC 13344668 & TIC 63921468 & Unit \\
    \hline
    \multicolumn{6}{c}{\textit{Fitted parameters}} \\ 
    \hline 
    $R_b / R_\star$ & $0.0792\pm0.0027$ & $0.1376_{-0.0065}^{+0.0052}$ & $0.1414_{-0.0068}^{+0.0059}$ & $0.03626_{-0.00073}^{+0.00087}$ & \\ 
    $(R_\star + R_b) / a_b$ & $0.0652_{-0.0022}^{+0.0028}$ & $0.00683_{-0.00023}^{+0.00027}$ & $0.00738_{-0.00023}^{+0.00026}$ & $0.01920_{-0.00043}^{+0.00047}$ & \\ 
    $\cos{i_b}$ & $0.0183_{-0.011}^{+0.0097}$ & $0.0055_{-0.0018}^{+0.0012}$ & $0.0085_{-0.0046}^{+0.0022}$ & $0.01633\pm0.00079$ & \\ 
    $T_{0;b}^a$ & $2163.6472_{-0.0012}^{+0.0012}$ & $1350.1844_{-0.0016}^{+0.0017}$ & $1445.0473_{-0.0010}^{+0.0015}$ & $1400.5143_{-0.0087}^{+0.0034}$ & $\mathrm{BJD}$ \\ 
    $P_b$ & $8.5821566_{-0.0000072}^{+0.0000072}$ & $175.532\pm0.097$ & $141.77_{-0.63}^{+0.72}$ & $131.331\pm0.010$ & $\mathrm{d}$ \\ 
    $K_b$ & $1.915\pm0.030$ & $0.862\pm0.027$ & $4.8_{-1.9}^{+2.5}$ & $2.0222_{-0.011}^{+0.0099}$ & $\mathrm{km\,s^{-1}}$ \\ 
    $\sqrt{e_b} \cos{\omega_b}$ & $0.035_{-0.12}^{+0.090}$ & $-0.228\pm0.040$ & $-0.65_{-0.11}^{+0.17}$ & $-0.2723_{-0.0088}^{+0.0095}$ & \\ 
    $\sqrt{e_b} \sin{\omega_b}$ & $0.134_{-0.10}^{+0.060}$ & $0.676_{-0.022}^{+0.020}$ & $0.551_{-0.092}^{+0.084}$ & $0.3213\pm0.0084$ & \\ 
    $q_{1; \mathrm{TESS}}$ & $0.29_{-0.21}^{+0.39}$ & $0.62_{-0.29}^{+0.26}$ & $0.35_{-0.28}^{+0.41}$ & $0.19_{-0.12}^{+0.18}$ & \\ 
    $q_{2; \mathrm{TESS}}$ & $0.22_{-0.16}^{+0.33}$ & $0.48\pm0.28$ & $0.27_{-0.20}^{+0.35}$ & $0.43_{-0.30}^{+0.36}$ & \\  
    $\ln{\sigma_\mathrm{TESS}}$ & $-6.846_{-0.033}^{+0.035}$ & $-6.53_{-0.12}^{+0.13}$ & $-6.64_{-0.13}^{+0.15}$ & $-8.776\pm0.077$ & $\ln{ \mathrm{rel. flux.} }$ \\ 
    $\ln{\sigma_\mathrm{jitter; FEROS}}$ & $-2.66_{-0.28}^{+0.32}$ & $-2.42_{-0.13}^{+0.14}$ & $-1.70_{-0.17}^{+0.18}$ & $-4.19_{-4.5}^{+0.63}$ & $\ln{ \mathrm{km/s} }$ \\ 
    \\
    \hline
    \hline
    \end{tabular}
    \tablefoot{a -- BJD - 2,457,000}
\end{table*} 

\begin{table*}
 \centering
 \caption{Derived parameters from the {\tt allesfitter} analysis.} \label{tab:sec_der}
 \begin{adjustbox}{width=0.99\textwidth}
 \begin{tabular}{lcccc}
    \hline
    \hline
    Parameter & TIC 9344899 & TIC 52059926 & TIC 13344668 & TIC 63921468 \\
    \hline
    \multicolumn{5}{c}{\textit{Derived parameters}} \\ 
    \hline
    Host radius over semi-major axis; $R_\star/a_\mathrm{b}$ & $0.0604_{-0.0020}^{+0.0026}$ & $0.00601_{-0.00019}^{+0.00023}$ & $0.00647_{-0.00020}^{+0.00023}$ & $0.01853_{-0.00042}^{+0.00045}$\\ 
    Semi-major axis over host radius; $a_\mathrm{b}/R_\star$ & $16.55_{-0.67}^{+0.57}$ & $166.4_{-6.0}^{+5.4}$ & $154.6_{-5.3}^{+4.9}$ & $54.0\pm1.3$\\ 
    Companion radius over semi-major axis b; $R_\mathrm{b}/a_\mathrm{b}$ & $0.00478_{-0.00022}^{+0.00029}$ & $0.000825\pm0.000054$ & $0.000913_{-0.000051}^{+0.000056}$ & $0.000672_{-0.000022}^{+0.000025}$\\ 
    Companion radius; $R_\mathrm{b}$ ($\mathrm{R_{\oplus}}$) & $9.16\pm0.40$ & $10.71_{-0.57}^{+0.52}$ & $10.32\pm0.57$ & $9.30_{-0.27}^{+0.30}$\\ 
    Companion radius; $R_\mathrm{b}$ ($\mathrm{R_{jup}}$) & $0.817\pm0.036$ & $0.955_{-0.051}^{+0.047}$ & $0.921\pm0.051$ & $0.830_{-0.024}^{+0.027}$\\ 
    Semi-major axis; $a_\mathrm{b}$ ($\mathrm{R_{\odot}}$) & $17.52_{-0.86}^{+0.80}$ & $118.8\pm5.4$ & $103.4\pm4.6$ & $126.7\pm4.1$\\ 
    Semi-major axis; $a_\mathrm{b}$ (AU) & $0.0815_{-0.0040}^{+0.0037}$ & $0.553\pm0.025$ & $0.481\pm0.022$ & $0.589\pm0.019$\\ 
    Inclination; $i_\mathrm{b}$ (deg) & $88.95_{-0.56}^{+0.63}$ & $89.685_{-0.071}^{+0.11}$ & $89.51_{-0.13}^{+0.27}$ & $89.064\pm0.046$\\ 
    Eccentricity; $e_\mathrm{b}$ & $0.029_{-0.018}^{+0.021}$ & $0.510\pm0.026$ & $0.747_{-0.13}^{+0.062}$ & $0.1773\pm0.0051$\\ 
    Argument of periastron; $w_\mathrm{b}$ (deg) & $80._{-33}^{+72}$ & $108.6\pm3.4$ & $139.5_{-12}^{+9.5}$ & $130.3\pm1.5$\\ 
    Mass ratio; $q_\mathrm{b}$ & $0.01890_{-0.00090}^{+0.0010}$ & $0.0221_{-0.0012}^{+0.0013}$ & $0.094_{-0.028}^{+0.038}$ & $0.0425_{-0.0014}^{+0.0015}$\\ 
    Companion mass; $M_\mathrm{b}$ ($\mathrm{M_{\oplus}}$) & $6170_{-580}^{+610}$ & $5380_{-370}^{+390}$ & $21500_{-6500}^{+8800}$ & $21510\pm1000$\\ 
    Companion mass; $M_\mathrm{b}$ ($\mathrm{M_{jup}}$) & $19.4_{-1.8}^{+1.9}$ & $16.9_{-1.1}^{+1.2}$ & $68_{-20.}^{+28}$ & $67.7\pm3.3$\\ 
    Companion mass; $M_\mathrm{b}$ ($\mathrm{M_{\odot}}$) & $0.0185_{-0.0017}^{+0.0018}$ & $0.0162_{-0.0011}^{+0.0012}$ & $0.065_{-0.020}^{+0.026}$ & $0.0646\pm0.0031$\\ 
    Impact parameter; $b_\mathrm{tra;b}$ & $0.30_{-0.18}^{+0.15}$ & $0.460_{-0.16}^{+0.100}$ & $0.41\pm0.25$ & $0.750_{-0.021}^{+0.023}$\\ 
    Total transit duration; $T_\mathrm{tot;b}$ (h) & $4.00_{-0.10}^{+0.12}$ & $4.87_{-0.14}^{+0.16}$ & $3.36_{-0.22}^{+0.28}$ & $11.56_{-0.34}^{+0.20}$\\  
    Host density from orbit; $\rho_\mathrm{\star;b}$ (cgs) & $1.16_{-0.14}^{+0.12}$ & $2.83\pm0.30$ & $3.47\pm0.34$ & $0.172\pm0.012$\\ 
    Companion density; $\rho_\mathrm{b}$ (cgs) & $44.1_{-7.0}^{+8.4}$ & $24.1_{-3.6}^{+4.8}$ & $111_{-42}^{+51}$ & $147_{-16}^{+18}$\\ 
    Companion surface gravity; $g_\mathrm{b}$ (cgs) & $71000_{-8000}^{+7200}$ & $45100_{-5400}^{+6500}$ & $199000_{-71000}^{+76000}$ & $244000\pm17000$\\ 
    Equilibrium temperature; $T_\mathrm{eq;b}$ (K)$^1$ & $923_{-19}^{+22}$ & $229.6_{-5.1}^{+5.5}$ & $237.7_{-5.2}^{+5.6}$ & $572.4\pm9.3$\\ 
    Transit depth (undil.); $\delta_\mathrm{tr; undil; b; TESS}$ (ppt) & $7.10_{-0.34}^{+0.36}$ & $22.94_{-0.81}^{+0.90}$ & $22.50_{-0.86}^{+0.82}$ & $1.316_{-0.049}^{+0.055}$\\ 
    Transit depth (dil.); $\delta_\mathrm{tr; dil; b; TESS}$ (ppt) & $3.52_{-0.17}^{+0.18}$ & $22.94_{-0.81}^{+0.90}$ & $22.50_{-0.86}^{+0.82}$ & $1.316_{-0.049}^{+0.055}$\\ 
    \hline
    \hline
 \end{tabular}
 \end{adjustbox}
\tablefoot{1 -- computed assuming zero albedo}
\end{table*}

\begin{figure*}
\centering
\includegraphics[width=0.83\textwidth, trim= {0.0cm 0.0cm 0.0cm 0.0cm}]{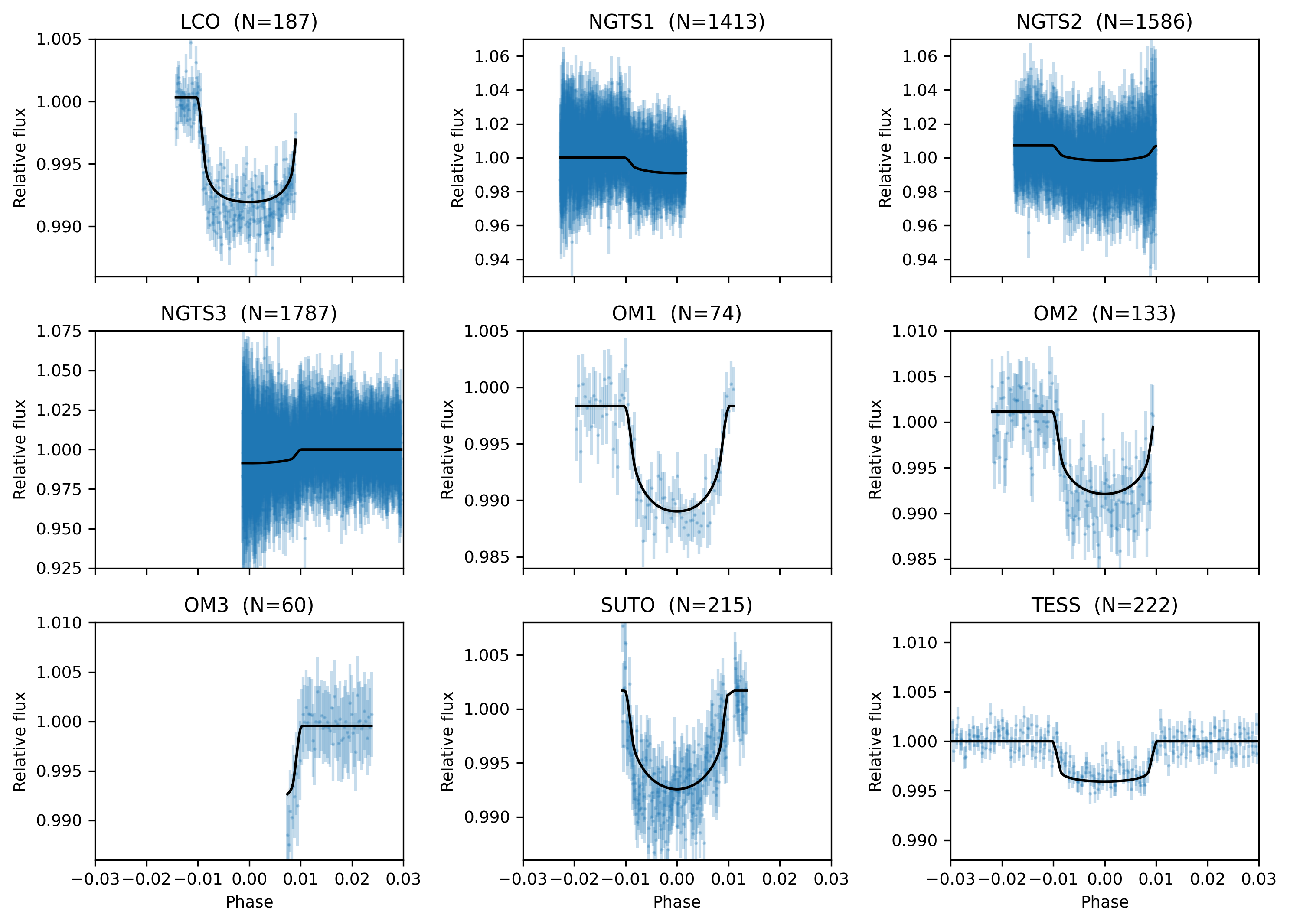}
\caption{All phased LCs for TIC\,9344899. The gray points represent TESS data, and the blue points are TESS binned data in the phase curve with a bin width of 24 minutes in phase. The red lines represent the best transit models.} \label{fig:transit_all}
\end{figure*}

\begin{figure*}
\centering
\includegraphics[width=0.45\textwidth, trim= {0.0cm 0.0cm 0.0cm 0.0cm}]{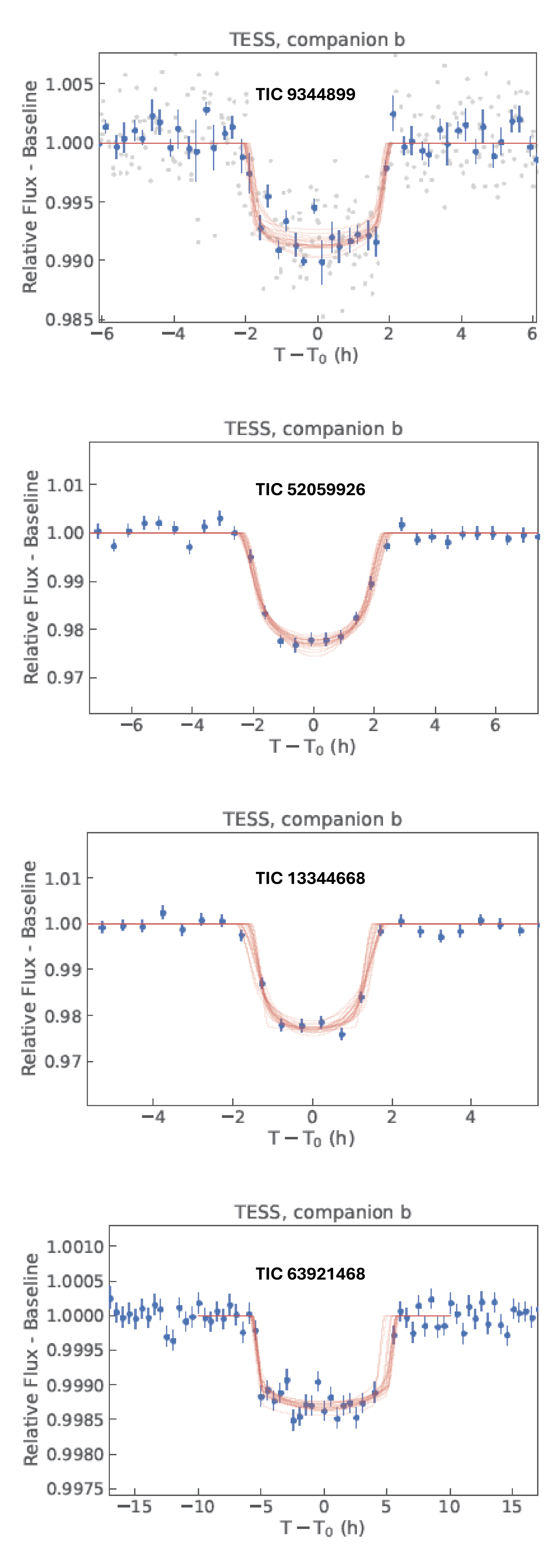}
\caption{The phased LCs fitted with {\tt allesfitter} as described in Section \ref{sec:RVs}. The gray points represent TESS data, and the blue points are TESS binned data in the phase curve with a bin width of 24 minutes in phase. The red lines represent the best transit models.} \label{fig:transit}
\end{figure*}

\begin{figure}[!htbp]
\centering
  \includegraphics[width=13cm]{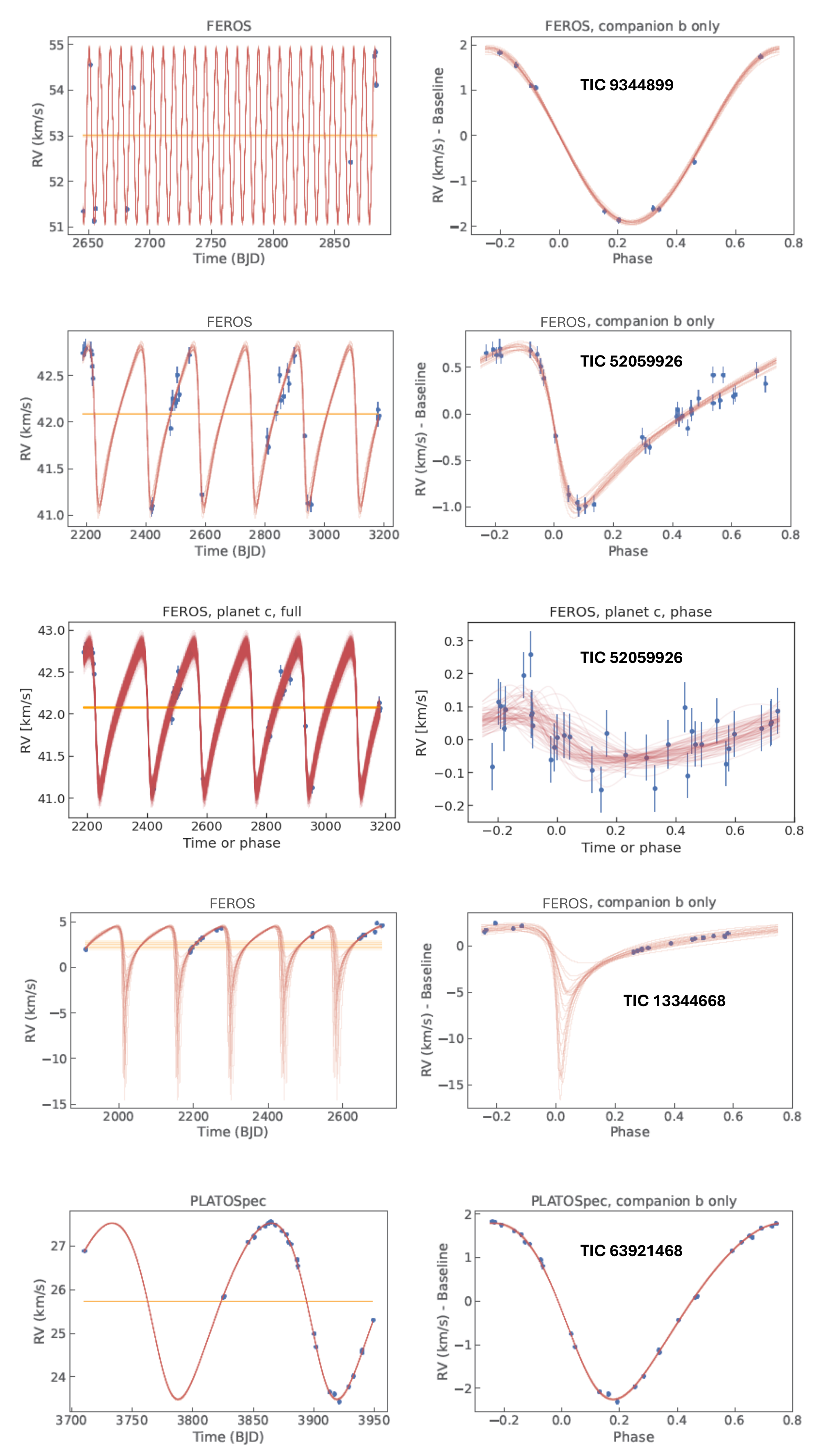}
     \caption{The radial velocity curves fitted with the {\tt allesfitter} as described in Section \ref{sec:RVs}. The red lines represent the best RV models.}
     \label{fig:RVs}
\end{figure}

\end{appendix}

\end{document}